\documentclass[prb,twocolumn,aps,superscriptaddress]{revtex4-1}
\usepackage{graphicx}
\usepackage{dcolumn}
\usepackage{bm}
\usepackage{amssymb}
\usepackage{amsmath}
\usepackage{subfigure}
\usepackage{physics}
\usepackage{sublabel}
\usepackage[utf8]{inputenc}

\usepackage{hyperref}

\begin{document}

\title{Proposal for a graphene based nano-electro-mechanical reference piezoresistor}

\author{Abhinaba Sinha}
\affiliation{Department of Electrical Engineering, IIT Bombay, Powai, Mumbai-400076, India}
	
\author{Abhishek Sharma}
\affiliation{Department of Electrical Engineering, IIT Ropar, Rupnagar-140001, India}

\author{Pankaj Priyadarshi}
\affiliation{Department of Electrical Engineering, IIT Bombay, Powai, Mumbai-400076, India}
	
\author{Ashwin Tulapurkar}
\affiliation{Department of Electrical Engineering, IIT Bombay, Powai, Mumbai-400076, India}
	
\author{Bhaskaran Muralidharan}
\affiliation{Department of Electrical Engineering, IIT Bombay, Powai, Mumbai-400076, India}
	
\begin{abstract}
Motivated by the recent prediction of anisotropy in piezoresistance of ballistic graphene along longitudinal and transverse directions, we investigate the angular gauge factor of graphene in the ballistic and diffusive regimes using highly efficient quantum transport models. It is shown that the angular guage factor in both ballistic and diffusive graphene between $0^{\circ}$ to $90^{\circ}$ bears a sinusoidal relation with a periodicity of $\pi$ due to the reduction of six-fold symmetry into a two-fold symmetry as a result of applied strain. The angular gauge factor is zero at critical angles $20^{\circ}$ and $56^{\circ}$ in ballistic and diffusive regimes respectively. Based on these findings, we propose a graphene based ballistic nano-sensor which can be used as a reference piezoresistor in a Wheatstone bridge read-out technique. The reference sensors proposed here are unsusceptible to inherent residual strain present in strain sensors and unwanted strain generated by the vapours in explosives detection. The theoretical models developed in this paper can be applied to explore similar applications in other 2D-Dirac materials. The proposals made here potentially pave the way for implementation of NEMS strain sensors based on the principle of ballistic transport, which will eventually replace MEMS piezoresistance sensors with a decrease in feature size. The presence of strain insenstive ``critical angle'' in graphene may be useful in flexible wearable electronics also.
\end{abstract}

\maketitle
\section{Introduction}
The development of micro/nano-electromechanical systems (MEMS/NEMS) has brought significant changes in every aspect of human life. The applications of MEMS in areas such as biotechnology~\cite{Madou2003,Wang2006}, medicine~\cite{Tsai2007,Kleinstreuer2008}, avionics~\cite{Jang2007,Leclerc2007}, particle transportation~\cite{Desai1999}, and defense~\cite{Tang2001,Siepmann2006} are virtually limitless. High-performance micro-scale systems, devices, and structures, including transducers~\cite{Nam2005,Mcneil2005}, switches~\cite{Rebeiz2001,Brown1998}, logic gates~\cite{Tsai2008,Tsai2010}, actuators~\cite{Bell2005,Conway2007} and sensors~\cite{Reeds2003,Kitano2006} are currently used in day to day life. Graphene, a single atom thick material possesses extraordinary electro-mechanical properties such as high elasticity~($\approx20\%$)~\cite{Liu2007,Kim2009}, Young's modulus~($\approx1$~TPa)~\cite{Lee2008}, mobility~\cite{Novoselov2004} and mean free path (in sub-micron range)~\cite{Novoselov2004,Adam2008,CastroNeto2009}.
Due to these properties, it is considered a promising material for next generation micro/nano electro-mechanical systems. Graphene is already used in MEMS systems as sensors~\cite{Smith2013,Dolleman2015,Wang2016}, switches~\cite{Li2018}, resonators~\cite{Chen2009} and actuators~\cite{Huang2012,Rogers2011}, to name a few. \\
\indent Rapid miniaturization of MEMS systems as a result of state-of-the-art nano-fabrication techniques, on one hand, offers multiple applications in a single chip, and on the other, necessitates the revamping of the theoretical understanding of electronic transport processes at a microscopic level in both the ballistic \cite{supriyo2012lessons,Urvesh,Priyadarshi,Swarnadip,Lodha} and the diffusive regimes \cite{Aniket,Aniket_2}. A deeper theoretical understanding of electronic transport across these systems will hence lead to novel functionalities that govern the next generation NEMS devices. In this work, we explore the use of graphene in strain sensing in both the ballisic and diffusive limit.\\
\indent The piezoresistance of graphene in both ballistic and diffusive regimes has been studied previously by various groups~\cite{Sinha2019,Huang2011,Smith2013,Smith2016}. The value of the gauge factor(GF) of ballistic and diffusive graphene for a uniaxial strain is reported in the range 0.3-6.1~\cite{Sinha2019,Smith2013,Smith2016}. In ballistic graphene, an anisotropy of the order of ten exists between the longituginal gauge factor (LGF) and the transverse gauge factor(TGF)~\cite{Sinha2019}. Motivated by the presence of such an anisotropy in GF, we further venture to explore the variation of GF along different directions ($0^{\circ}$ to $90^{\circ}$). \\
\indent We devise a theoretical model using quantum transport theory built from the tight binding representation to calculate the angular gauge factor (AGF) in the ballistic regime. \\
Our model is highly efficient and thus reduces the computation time to $3\%$ of that required by the conventional band counting method in Ref.~\cite{Sinha2019}. The theoretical model used in this paper can be extended to other 2D-Dirac materials \cite{Ales_1,Ales_2} as well. We obtain the AGF in the diffusive regime using the conductivity model developed by Peres~\textit{et.~al}~\cite{Peres2007}. The value of GF simulated previously in the diffusive regime uses an approximation for Fermi-velocity instead of the actual value~\cite{Huang2011,Sinha2019}. In this work, we calculate the actual value of Fermi-velocity along different directions which enables us to get an accurate value of AGF in the diffusive regime. We find that the AGF in ballistic and diffusive graphene is a sinusoidal function of the transport direction with a periodicity of $\pi$ due to a reduction of the six-fold symmetry into two-fold symmetry on the application of a uniaxial strain. The AGF becomes zero at the critical angles $20^{\circ}$ and $56^{\circ}$ in ballistic and diffusive graphene respectively. Using these results, we propose a ballistic nano-sensor and a reference resistor using graphene in a Wheatstone bridge based read-out technique. Further, the proposals made here potentially pave the way for implementation of NEMS strain sensors based on the principle of ballistic transport which in NEMS sensors with further decrease in feature size. The presence of strain insenstive ``critical angle'' in graphene may be useful in flexible wearable electronics also. \\
\indent In the subsequent sections, we develop the mathematical model to calculate the AGF of graphene across different transport regimes, explain the underlying physics for the predicted results, and discuss the applications and future scopes. The detail derivation of mathematical expressions are given in Appendix.
\section{Theoretical Model}
\subsection{Simulation setup}
\begin{figure}
	
	\subfigure[]{\includegraphics[height=0.2\textwidth,width=0.228\textwidth]{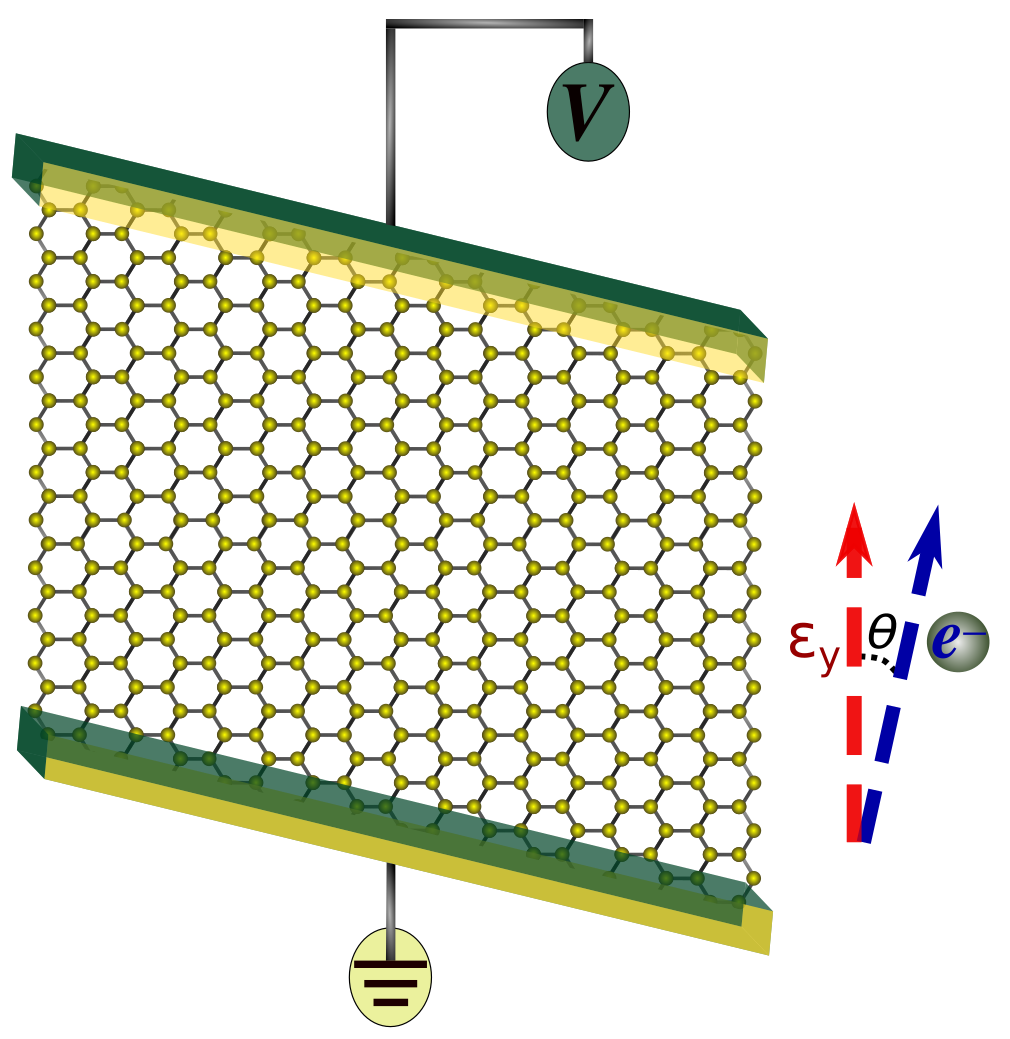}\label{PO2_Setup}}
	\quad
	\subfigure[]{\includegraphics[height=0.14\textwidth,width=0.228\textwidth]{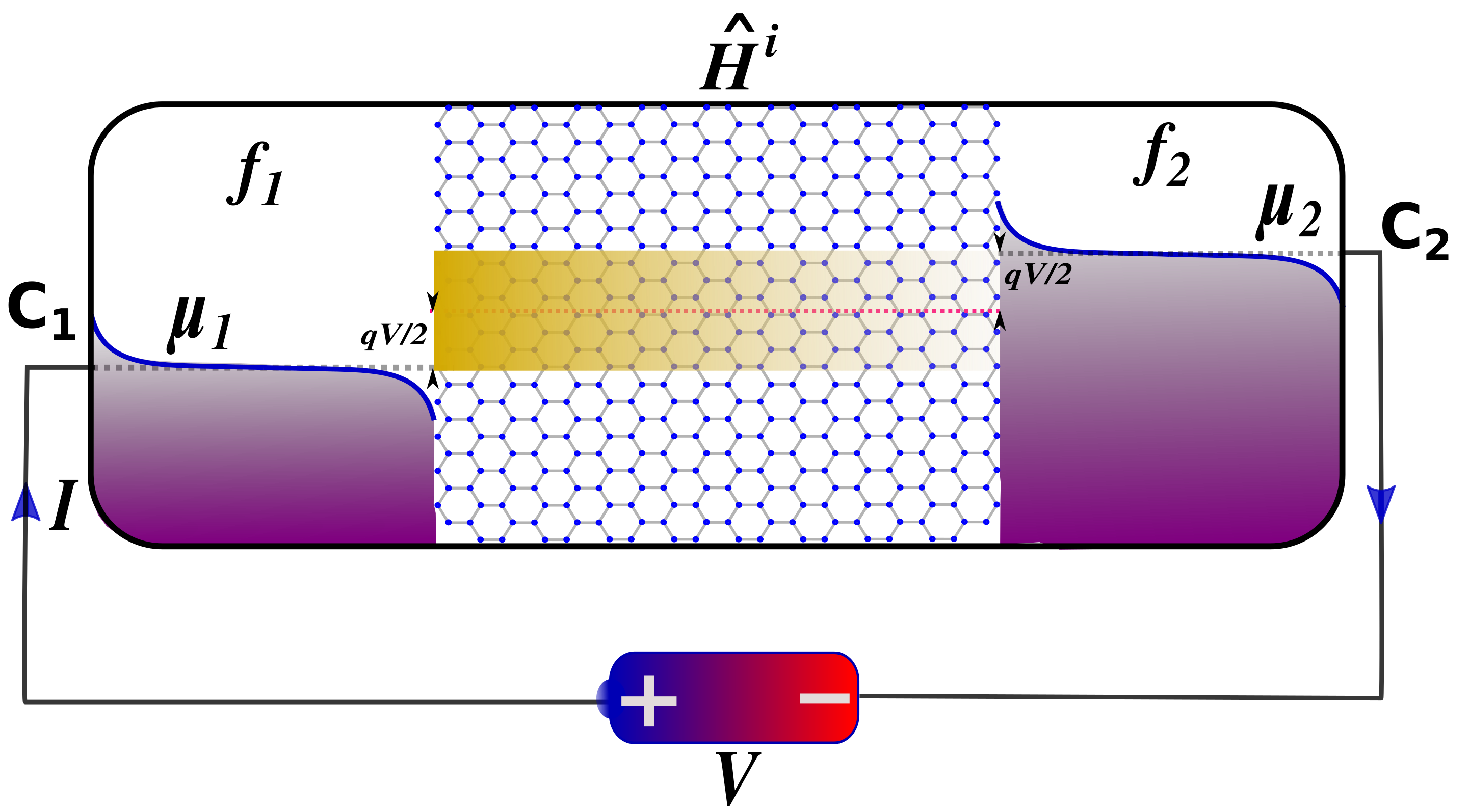}\label{P02_QTM}}
	\quad
	\caption{ Description of device schematic and transport setup for AFG calculation. (a) Schematic of a uniaxial strained graphene along the zigzag direction (y axis). The red and blue arrows  represent the respective directions of applied strain and electron transport. The angle $\theta$ between them varies  within the limit (0,$\pi/2$). (b) Voltage driven charge transport model across a strained graphene sheet (with Hamiltonian $\hat {H^{i}}$, where $i$ denotes strain percentage) sandwiched between electronic contacts $C_1$ and $C_2$ with Fermi-levels $\mu_1=\mu_0-qV/2$ and $\mu_2=\mu_0+qV/2$  respectively. Transport formalism is restricted within the linear response regime.}
	\label{P02_Fig_1}
\end{figure}

The schematic diagram of the angular gauge factor setup is presented in Fig.~\ref{PO2_Setup}. A uniaxial strain ($\varepsilon_{y}$) is applied along the zigzag direction, and the resistance is computed along different directions represented by `$\theta$'. \\
\indent The quantum transport model for the given setup is schematized in Fig.~\ref{P02_QTM}. The setup constitutes a graphene sheet represented by  Hamiltonian `$\hat{H^{i}}$' and ideal reflection-less contacts $C_1$ and $C_2$. The equilibrium Fermi-energy of the graphene sheet and contacts are maintained at 0 eV. The voltage applied at the terminals $C_{1}$ and $C_{2}$ are $\textrm{-V/2}$ and $\textrm{V/2}$ Volts respectively.\\
\indent We note that the simulation setup described here evaluates AGF in the linear regime [-0.01 eV - 0.01eV] within the elastic limit of graphene [$0\%-10\%$].\\

\subsection{Angular gauge factor calculation}
The transport properties of graphene in the ballistic regime depends on mode density~\cite{supriyo2012lessons} whereas in the diffusive regime depends on Fermi velocity~\cite{Peres2007}. The value of mode density and Fermi-velocity depend on the applied strain. We evaluate the mode density and Fermi velocity of graphene as a function of strain along different directions ($\theta$) from band-structure of strained graphene. The mathematical models derived using quantum transport, and semi-classical transport formalisms for the evaluation of AGF in different transport regimes of graphene are discussed further. 
\subsubsection{Tight binding model}
The tight binding Hamiltonian of a honey-comb lattice is expressed as
\begin{equation}
\hat{H^{i}}= \sum_{l,\tau}^{}t^{i}_{\tau}c^{}_{l}c^{\dagger}_{\tau} + \mathrm{H.c.}
\label{P02_eqH}
\end{equation}
In Eq.~\eqref{P02_eqH}, $t^{i}_{\tau}$ represents the hopping parameter that connects the lattice site $l$ with its neighbours $\tau$ in graphene at strain $\varepsilon_{y}=i\%$. $c^{}_{l}$ and $c^{\dagger}_{\tau}$ are respectively the annihilation and creation operators of electrons at sites $l$ and $\tau$. We consider that the electron dynamics of graphene is governed by the nearest neighbour tight binding Hamiltonian. Thus, the energy eigen-values of Eq.~\eqref{P02_eqH} is given by
\begin{equation}
E^{i}(k) = \pm \mid{t_1^{i}e^{-j\vec{k}\cdot{\vec{ a_1^{i}}}} + t_2^{i} + t_3^{i}e^{-j\vec{k}\cdot{\vec{ a_2^{i}}}} }\mid,
\label{P02_eqEk}
\end{equation}
where $\vec{a_1^{i}}$ and $\vec{a_2^{i}}$ are the basis vectors of strained graphene, and $t_1^{i}$, $t_2^{i}$ and $t_3^{i}$ are the nearest neighbor hopping parameters.\\
\indent We obtain the tight-binding parameters of uniaxially strained graphene from Ribeiro \textit{et al.}~\cite{Ribeiro2009}. In Ref.~\cite{Ribeiro2009} the parameters are obtained by fitting Eq.~\ref{P02_eqEk} with \textit{ab-initio} band-structures. This model is valid for energy E in the range [-0.2 eV- 0.2 eV].\\
\indent The nearest neighbor tight binding model of graphene described by Eq.~\eqref{P02_eqEk} accurately predicts the shift in Dirac cones due to strain, band gap threshold and anisotropy in Fermi velocity~\cite{Hasegawa2006,Pereira2009}. This model is consistent with the \textit{ab-initio} calculations~\cite{Farjam2009,Ribeiro2009,Choi2010,Ni2009,Huang2011} and experiments~\cite{Kim2009,Huang2010}. Thus, Eq.~\eqref{P02_eqEk} is suitable for AGF calculation in the linear regime.\\
\subsubsection{Ballistic Regime}
\begin{figure}
	\centering	
	\subfigure[]{\includegraphics[height=0.2\textwidth,width=0.228\textwidth]{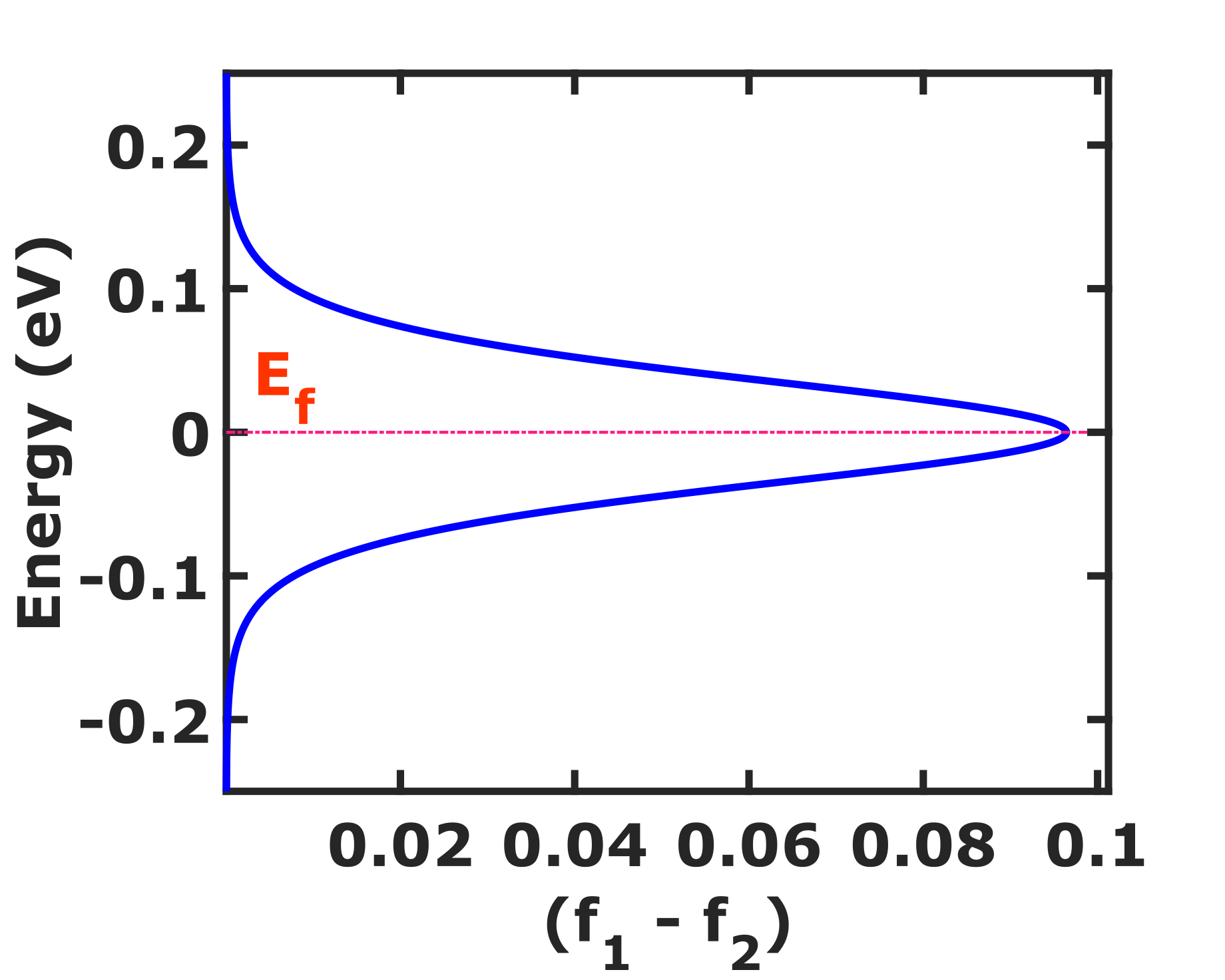}\label{P02_fermi_window}}
	\quad
	\subfigure[]{\includegraphics[height=0.2\textwidth,width=0.228\textwidth]{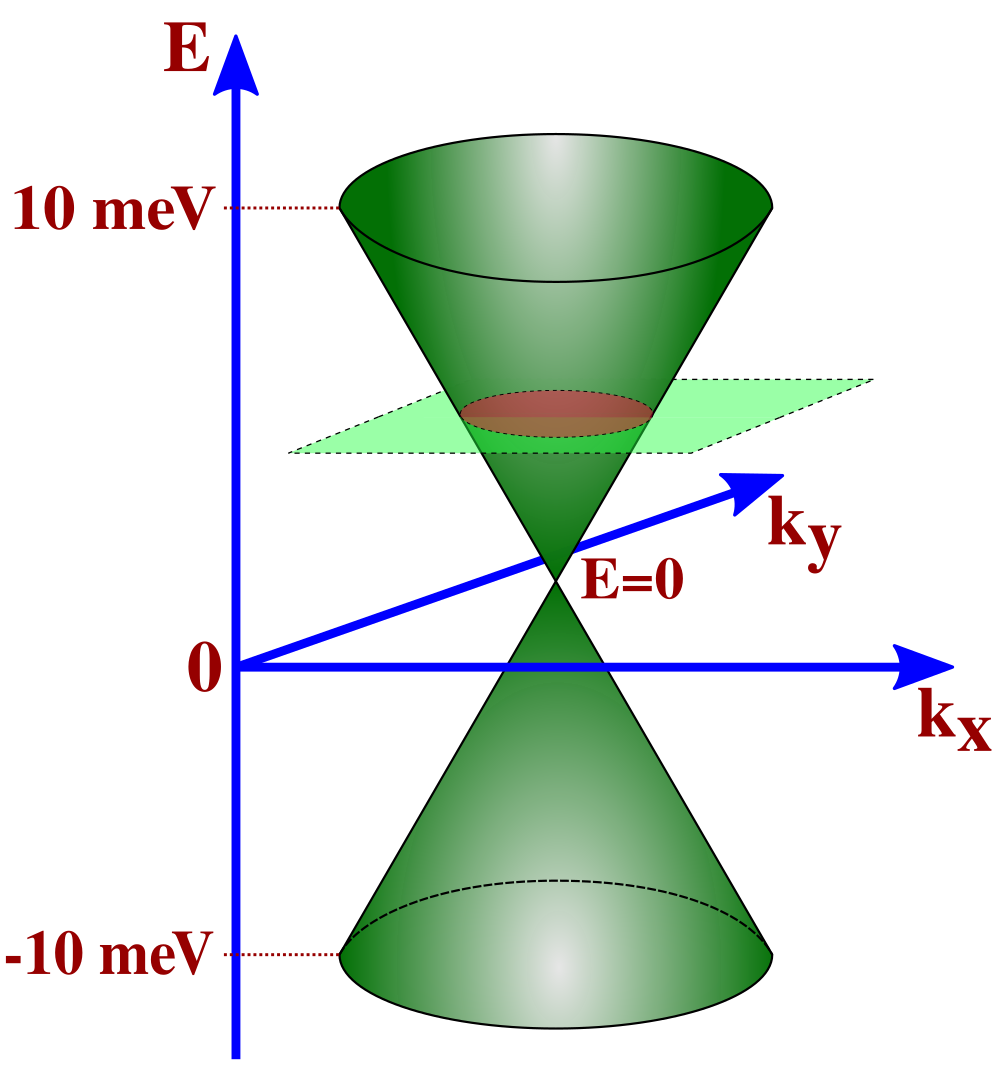}\label{P02_DC}}
	\quad
	\subfigure[]{\includegraphics[height=0.2\textwidth,width=0.228\textwidth]{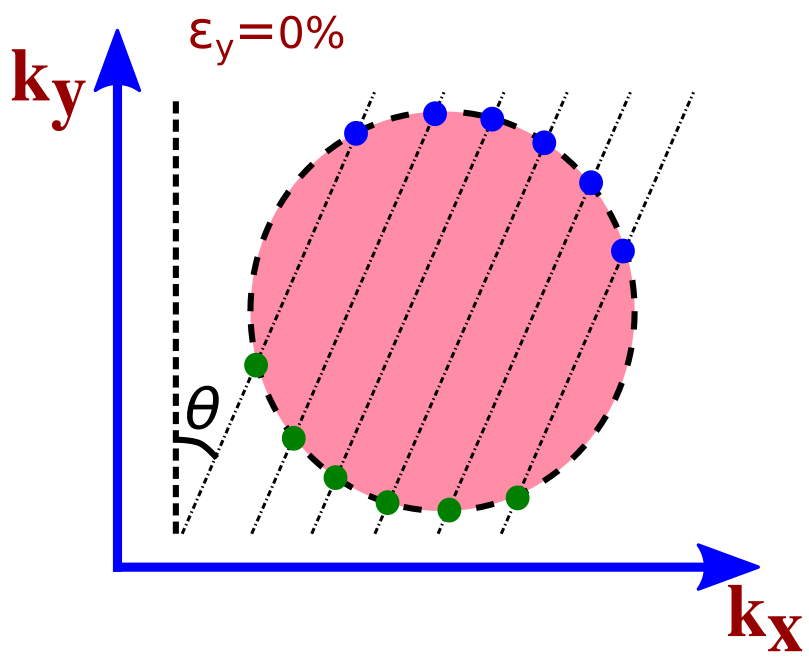}\label{P02_DC_circle}}	
	\quad
	\subfigure[]{\includegraphics[height=0.2\textwidth,width=0.228\textwidth]{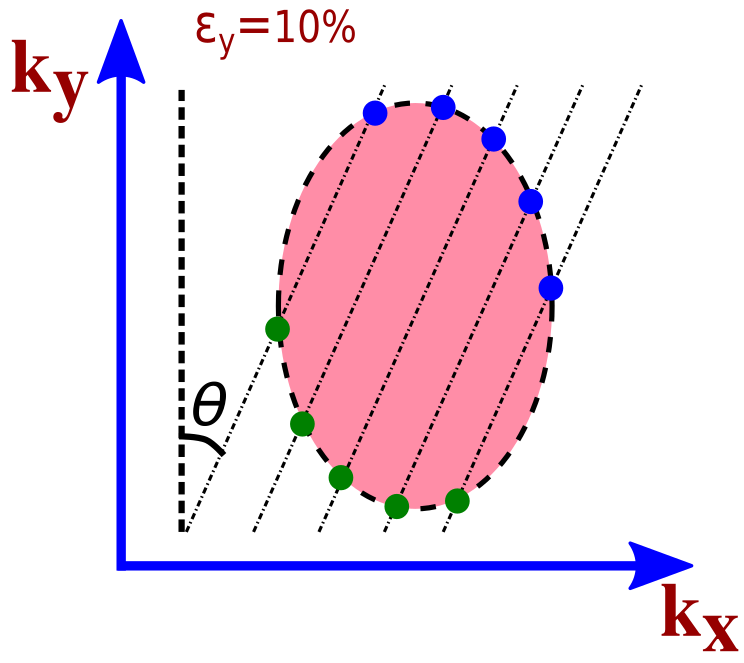}\label{P02_DC_ellipse}}
	\quad		
	\caption{ Validation of the tight binding model and mode density calculation in ballistic graphene. (a) Depiction of the Fermi-window $(f_1-f_2)$ as a function of energy predefined in Fig~1(b). The window opens up in the energy range [-0.2 eV,0.2 eV]. (b) 3-D view of graphene band-structure close to the Dirac point. (c-d) depict the constant energy surface and modes along $\theta$ . Constant energy surface shapes like (c) a circle for $0\%$ strain and like (d) an oval for $10\%$ strain respectively. In each case modes are depicted by blue dots. }	
	\label{sub-band}
\end{figure}

\begin{figure}
	\centering
	\subfigure[]{\includegraphics[height=0.225\textwidth,width=0.238\textwidth]{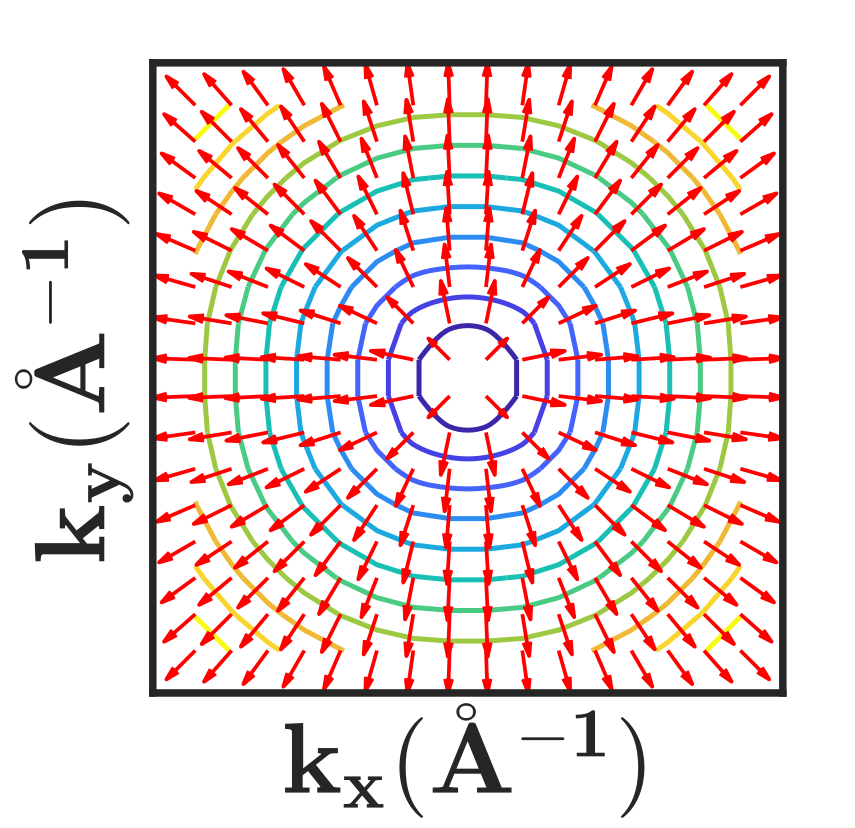}\label{P02_fermi_velocity_0}}
	\subfigure[]{\includegraphics[height=0.225\textwidth,width=0.238\textwidth]{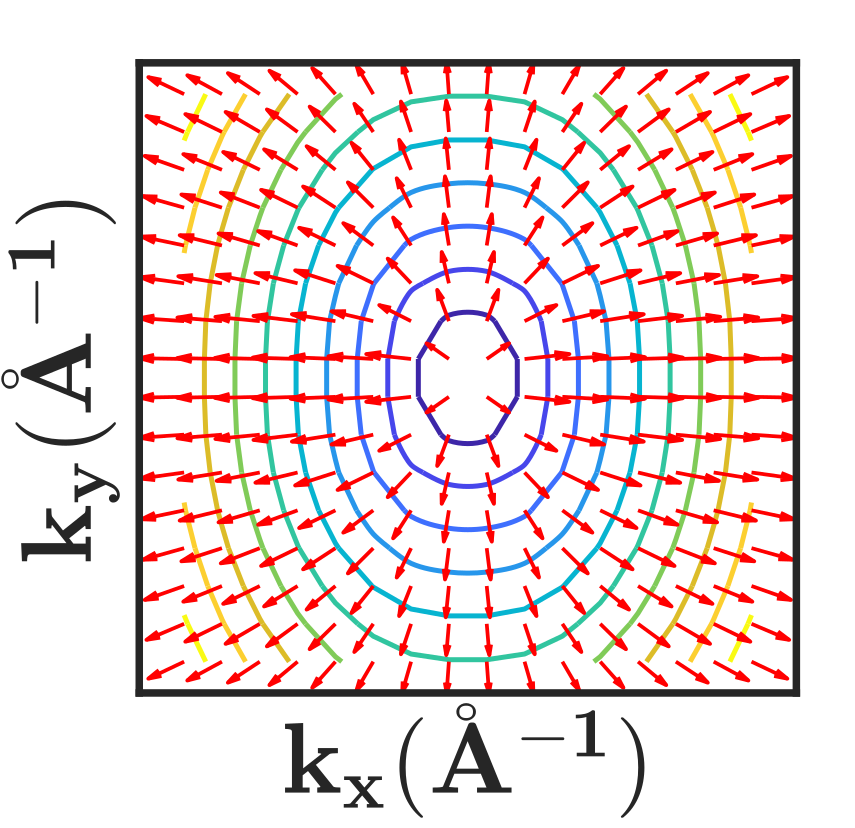}\label{P02_fermi_velocity_10}}
	\subfigure[]{\includegraphics[height=0.28\textwidth,width=0.228\textwidth]{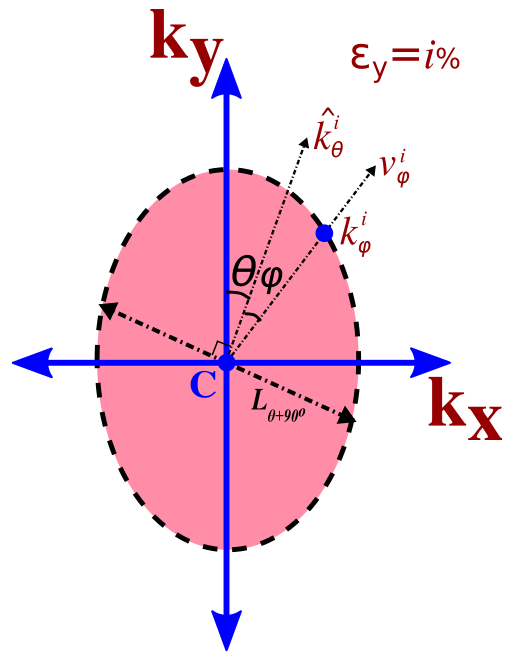}\label{P02_DC_fermi_velocity}}	
	\caption{ Calculation of average Fermi velocity in graphene for computing AGF in the diffusive regime. (a) and (b) depict the  Fermi velocity vectors (represented by the red arrows) and energy-contours near the Dirac-point for $\varepsilon_{y}=0 \% $ and $\varepsilon_{y}=10 \%$ respectively. (b) shows that strain induces anisotropy in Fermi velocity. (c) depicts the schematics of the unit vector $k^{i}_{\theta}$ along $\theta$ and Fermi velocity $v^{i}_{\phi}$  along  $\phi$.}
	\label{P02_Fermi-velocity}
\end{figure}

We compute the current-voltage characteristics of graphene in ballistic regime using Landauer formula which is expressed as
\begin{equation}
I^{i}_{\theta}(V)= \frac{2q}{h} \int_{-\infty}^{\infty} M^{i}_{\theta}(E)[f_{1}(E-{\mu_{1}})-f_{2}(E-{\mu_{2}})]dE,
\label{P02_eqI}	
\end{equation}
where $M^{i}_{\theta}(E)$ is the mode density at energy `$E$', strain-percentage `$i$' and electron transport direction `$\theta$'. The resistance calculated from Eq.~\eqref{P02_eqI} is expressed as
\begin{equation}
R^{i}_{\theta} = \dfrac{1}{d\{I^{i}_{\theta}(V)\}/dV}.
\end{equation} 
Thus, the value of AGF in the ballistic regime averaged over the entire strain-limit can be written as
\begin{subequations} \label{P02_eqbagf}
\begin{align}
\mathrm{(AGF)^{i}_{\theta}}= \bigg\{ \frac{R^{i}_{\theta}-R^{0}_{\theta}}{\varepsilon_{y}R^{0}_{\theta}}\bigg\},\\
(AGF)_{\theta} =\overline{(AGF)^{i}_{\theta}}. 
\end{align}
\end{subequations}
The Fermi-window $f_1(E-\mu_1)-f_2(E-\mu_2)$ depends on the energy and applied voltage. We apply a variable potential difference in the range [-0.01 V to 0.01 V] across the contacts. The Fermi-window plot as a function of energy at the potential difference  0.01 V is shown in Fig.~\ref{P02_fermi_window}. The Fermi-window opens up between -0.2 eV to 0.2 eV. Thus, our model using Eqs.~\eqref{P02_eqEk} and \eqref{P02_eqI} can accurately determine the angular piezo-resistance of graphene in the linear regime.\\
\indent We obtain the mode density `$M^{i}_{\theta}(E)$' from the band-structure in Dirac cone approximation. Figure~\ref{P02_DC} illustrates the constant energy surface formed as a result of intersection of the constant energy plane and the Dirac cone. Figures~\ref{P02_DC_circle} and \ref{P02_DC_ellipse} illustrate the constant energy surface and modes along $\theta$ at $0\%$ and $10\%$ strain. The blue and green dots represent the modes of forward and backward moving electrons, respectively. Since the effective number of Dirac cones within the first Brillouin zone of strained and unstrained graphene is two~\cite{Sinha2019}. Thus, the effective number of modes at energy `E' along $\theta$ is numerically equal to the sum of forward and backward moving electrons in a single Dirac cone. The mode density is therefore given by
\begin{equation}
M^{i}_{\theta}(E) = 2n^{i}_{\theta}(E),
\label{P02_eqmodes}
\end{equation}
where $n^{i}_{\theta}(E)$ is the number of transverse modes (TMs) intersecting the constant energy surface at energy `$E$'. The separation between TMs in Figs.~\ref{P02_DC_circle} and \ref{P02_DC_ellipse} is $2\pi/w^{i}_{\theta}$ where $w^{i}_{\theta}$ is the width of graphene sheet (see Appendix~\ref{P02_app1}). \\ 
\indent The theoretical model described above for the mode density calculation leads to a significant reduction in the computation-time compared to the full-band mode density calculation~\cite{Sinha2019} and can be applied to explore similar applications in other 2D-Dirac materials as well.

\begin{table}
	\caption{Variation of Fermi-velocity (in the scale of $\times10^5$m/s) with different values of strain ($\varepsilon_y$) and transport angle ($\theta$).}
	\centering
	\begin{tabular}{c@{\hskip 1.3 cm} c@{\hskip 1.3 cm} c@{\hskip 1.3 cm} c@{\hskip 1.3 cm} c} 
		\hline
		\hline						
		\large$\bm{\varepsilon_{y}}$ & $\bm{0^{\circ}}$ & $\bm{30^{\circ}}$ & $\bm{60^{\circ}}$ &  $\bm{90^{\circ}}$ \\		
		\hline			
		$\bm{0\%}$ & 8.41 & 8.41 & 8.41 & 8.41 \\
		$\bm{5\%}$ & 7.35 & 7.75 &8.37  & 8.59 \\
		$\bm{10\%}$ & 6.12 & 7.32 &8.45  & 8.77\\
		\hline
		\hline	
	\end{tabular}	
	\label{table:P02_FV}
\end{table}
\begin{figure}	
	\subfigure[]{\includegraphics[height=0.215\textwidth,width=0.225\textwidth]{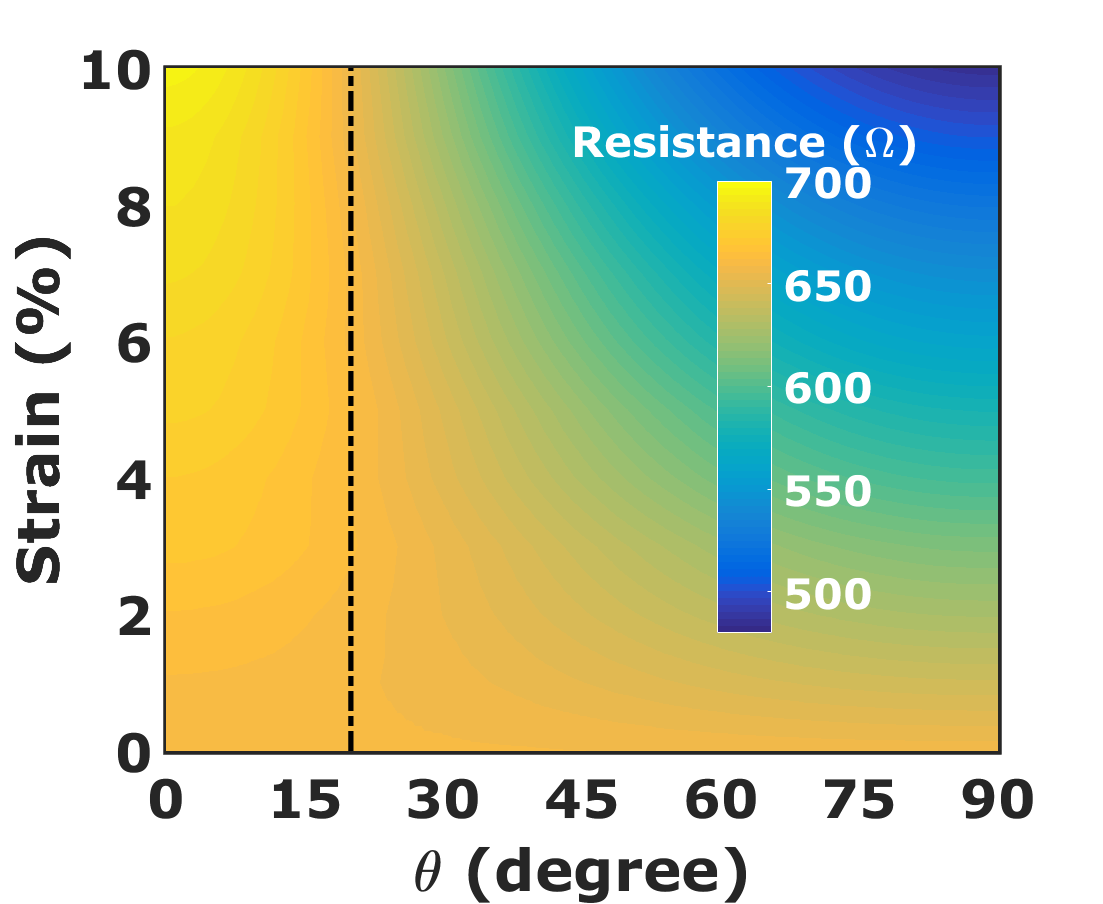}\label{P02_Resistance}}
	\quad
	\subfigure[]{\includegraphics[height=0.215\textwidth,width=0.225\textwidth]{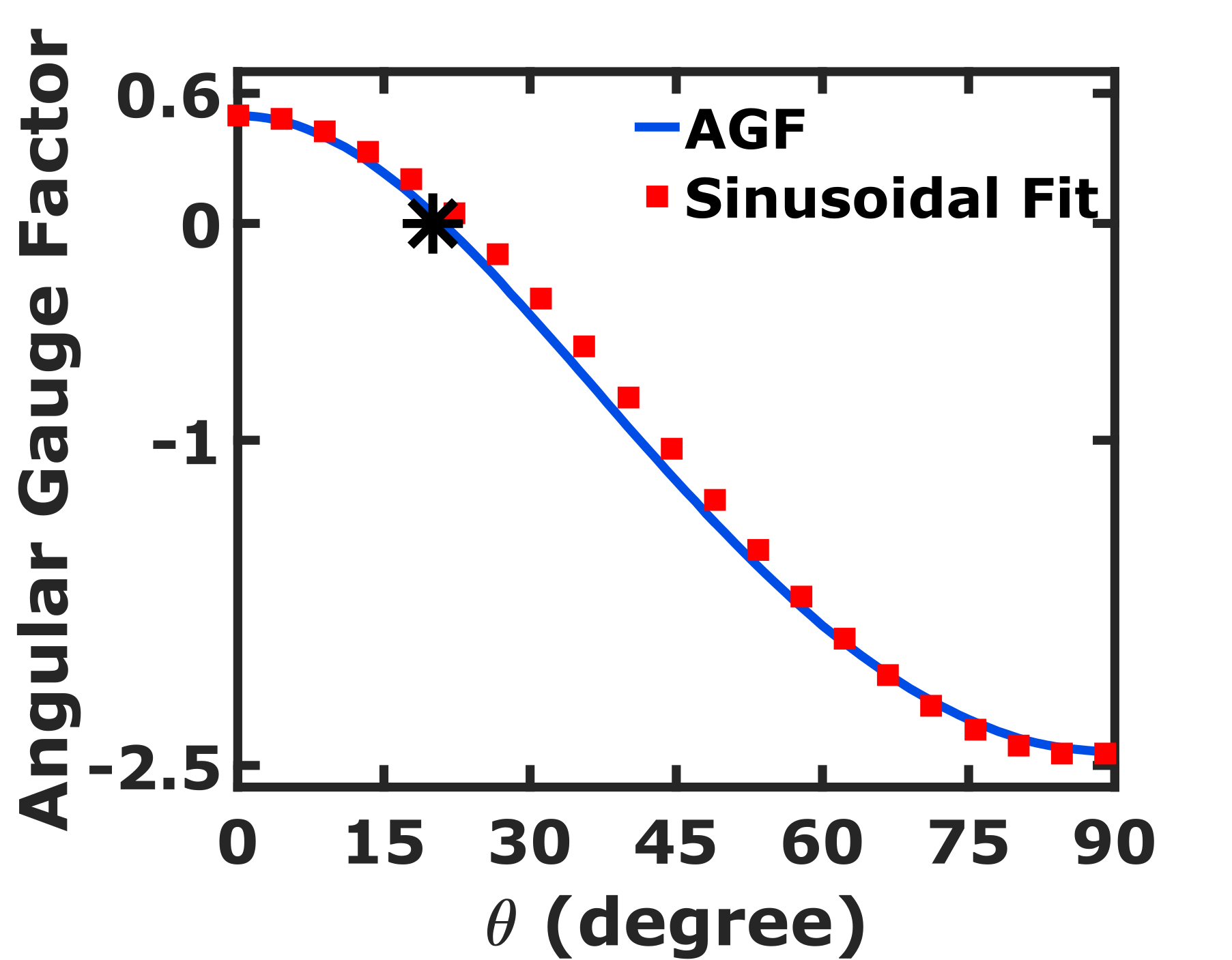}\label{P02_AGF_ballistic}}
	\caption{Piezoresistance along different transport directions in the ballistic graphene. (a) Color map of resistance of a 1-$\mu m$ wide graphene sheet as a function of $\varepsilon_y$ and $\theta$. (a) shows that the resistance remains constant at $\theta=20^{\circ}$. (b) depicts AGF along with its sinusoidal fit. It infers zero AGF at $\theta = 20^{\circ}$.}  		
	\label{Ballistic regime}
\end{figure}
\subsubsection{Diffusive Regime}
Electron transport in the diffusive regime is described via the semi-classical transport mechanism. The electrons are treated as classical particles (wave-packets) whose position and momentum are precisely known. The electron mobility determines the electron transport properties in the diffusive regime as compared to mode density in the ballistic regime. \\
\indent The experimentally determined conductivity of graphene in the diffusive regime shows a linear dependence on electron density except at the charge neutrality point~\cite{Novoselov2004}. By considering random Coulomb impurities as the dominant source of scattering, a linearly varying conductivity with gate voltage is obtained in Ref.~\cite{Nomura2007}. The expression for conductivity of graphene as a function of the electron density, considering the presence of charged impurities was formulated using Boltzman transport theory under the relaxation time approximation by Peres \textit{et al.} The expression for conductivity as derived in Ref.~\cite{Peres2007} is given by
\begin{equation}
\sigma = \frac{2e^{2}\pi (\hbar v_f)^2 n}{h {u^2_{o}}}.
\label{P02_eqc}
\end{equation}
\indent The conductivity of graphene in the diffusive regime depends on the electron density and Fermi velocity. However, variation in the conductivity due to a change in electron density is prominent only in the presence of a gate voltage~\cite{Pereira2009}.  In the absence of gate voltage, conductivity depends primarily on the Fermi-velocity. The anisotropy in resistance due to a tensile strain predicted using Eq.~\eqref{P02_eqc} comply with the experimental results~\cite{Kim2009}. Hence, we use Eq.~\eqref{P02_eqc} to obtain the expression for AGF in diffusive regime.\\
\indent The resistance of a uniaxially strained graphene is given by 
\begin{equation}
R^{i}_{\theta}=\frac{l^{i}_{\theta}}{\sigma^{i}_{\theta}~ w^{i}_{\theta}},
\label{P02_eqr}
\end{equation}
where $R^{i}_{\theta}$, $\sigma^{i}_{\theta}$, $l^{i}_{\theta}$, and $w^{i}_{\theta}$ are the resistance, conductivity, length and width respectively at strain `$i$' along the direction `$\theta$'.\\
\indent Thus, the expression for AGF in the diffusive regime averaged over the entire strain range is given by
\begin{subequations} \label{P02_eqdagf}
\begin{align}
\mathrm{(AGF)^{i}_{\theta}}=\frac{1}{\varepsilon_{y}}\bigg \{ \frac{\Delta l^{i}_{\theta}}{l^{0}_{\theta}}-\frac{\Delta w^{i}_{\theta}}{w^{0}_{\theta}} -2{\frac { \Delta v^{i}_{\theta}}{v^{0}_{\theta}}}\bigg\},\\
(AGF)_{\theta} =\overline{(AGF)^{i}_{\theta}}.
\end{align}
\end{subequations}
The AGF in diffusive regime depends on the variation of Fermi-velocity, electron density, and dimensions of the graphene with strain $\varepsilon_{y}$ and direction $\theta$. The strain-induced variation of Fermi-velocity is discussed in the present section, while the strain-induced change in dimensions is discussed separately in the next sub-section. \\
\indent  The velocity of electrons at an energy close to the Dirac point is equal to its Fermi-velocity. The Fermi-velocity is expressed as
\begin{equation}
	v^{i}_{\phi}=\frac{1}{\hbar}\{\nabla E^{i}({k})\}\bigg|_{k=k^{i}_{\phi}}.
	\label{P02_eqfv}
\end{equation}

Figures~\ref{P02_fermi_velocity_0} and \ref{P02_fermi_velocity_10} show the Fermi-velocity vectors and contours near a Dirac point at $0\%$ and $10\%$ strain respectively. The variation in magnitude of Fermi-velocity with $\varepsilon_{y}$ and $\theta$ is given in Table~\ref{table:P02_FV}. From the table, we infer that strain induces anisotropy in the Fermi-velocity. The AGF in diffusive regime depends on the average Fermi-velocity along the direction of transport. Figure~\ref{P02_DC_fermi_velocity} illustrates the methodology for evaluation of average Fermi-velocity along the transport direction ($\theta$). The average Fermi velocity along $\theta$ at strain `$i$' is expressed as
\begin{equation}
\overline{v^{i}_{\theta}}=\frac{1}{\pi} \int_{-\frac{\pi}{2}}^{\frac{\pi}{2}} \{v_{\phi}^{i}\cdot \hat{k^{i}_{\theta}}\} d\phi,
\label{P02_eqafv}
\end{equation}
where $v^{i}_{\phi}$ is the Fermi-velocity along direction $\phi$ and $\hat{k^{i}_{\theta}}$ is the unit vector along $\theta$. See Appendix~\ref{P02_app2} for detail derivation of Eq.~\eqref{P02_eqafv}.\\
\subsubsection{Strain distribution in graphene}
Apart from the variation of Fermi-velocity,  AGF also depends on the magnitude of strain along the transport ($\theta$) and transverse directions ($90^{\circ}+\theta$). The change in dimensions modifies the mode density and conductivity of graphene (see Eqs.~\eqref{P02_eqmodes} and \eqref{P02_eqdagf}).\\
\indent The strain `$\varepsilon_{y}$' generates components along different directions of graphene. The stiffness or compliance matrix due to a uniaxial strain along the basal plane of the graphene sheet is the same irrespective of the choice of coordinate axes~\cite{Landau1986}. Consequently, the Poisson's ratio of graphene sheet is the same irrespective of the direction of tensile strain in the basal plane~\cite{Pereira2009}.\\
\indent The mean free path of graphene is very high and is in the sub-micron range~\cite{Das2008, Novoselov2004, CastroNeto2009}. Therefore, we treat graphene as a continuum sheet in strain related calculations in this work.\\
\indent The strain components along the electron transport direction ($\varepsilon_{\theta}$) and its transverse direction ($\varepsilon^{\dagger}_{\theta}$) are expressed as
\begin{subequations} \label {P02_eq_ang_str}
\begin{align}	
\varepsilon_{\theta} &= \frac{1}{2}(\varepsilon_x + \varepsilon_y) + \frac{1}{2}(\varepsilon_y - \varepsilon_x)\cos2\theta,~~\mathrm{and} \label{P02_eq_str_theta} \\
\varepsilon^{\dagger}_{\theta} &= \frac{1}{2}(\varepsilon_x + \varepsilon_y) + \frac{1}{2}(\varepsilon_y - \varepsilon_x)\cos2(\theta+90^{\circ}) \label{P02_eq_str_90_theta},
\end{align}
\end{subequations}
where $\varepsilon_y$ is the longitudinal strain and $\varepsilon_x$ is the transverse strain ($-\sigma \varepsilon_{y}$). In ballistic regime, the mode density depends on the separation between TMs which is given by $2\pi/w^{i}_{\theta}$, where $w^{i}_{\theta}=w^{0}_{\theta}(1 + \varepsilon^{\dagger}_{\theta})$. In diffusive regime, apart from the Fermi-velocity, AGF varies with $\varepsilon_{\theta}$ and  $\varepsilon^{\dagger}_{\theta}$ (see Eq.~\ref{P02_eqdagf}).\\
\section{Results and Discussions } \label{P02_section3}
\begin{figure}[]
	\subfigure[]{\includegraphics[height=0.21\textwidth,width=0.229\textwidth]{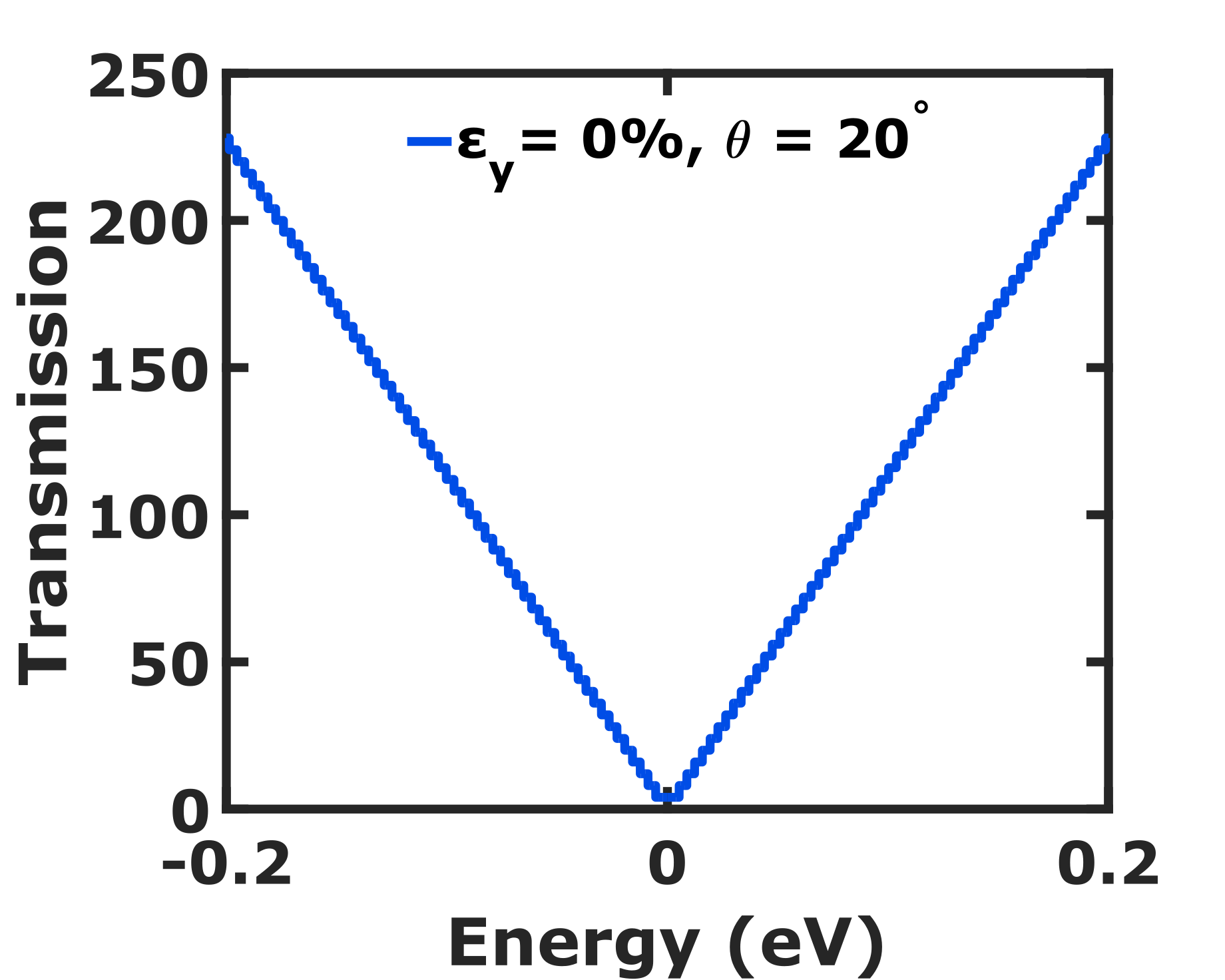}\label{P02_T_0_20}}
	\quad
	\subfigure[]{\includegraphics[height=0.21\textwidth,width=0.229\textwidth]{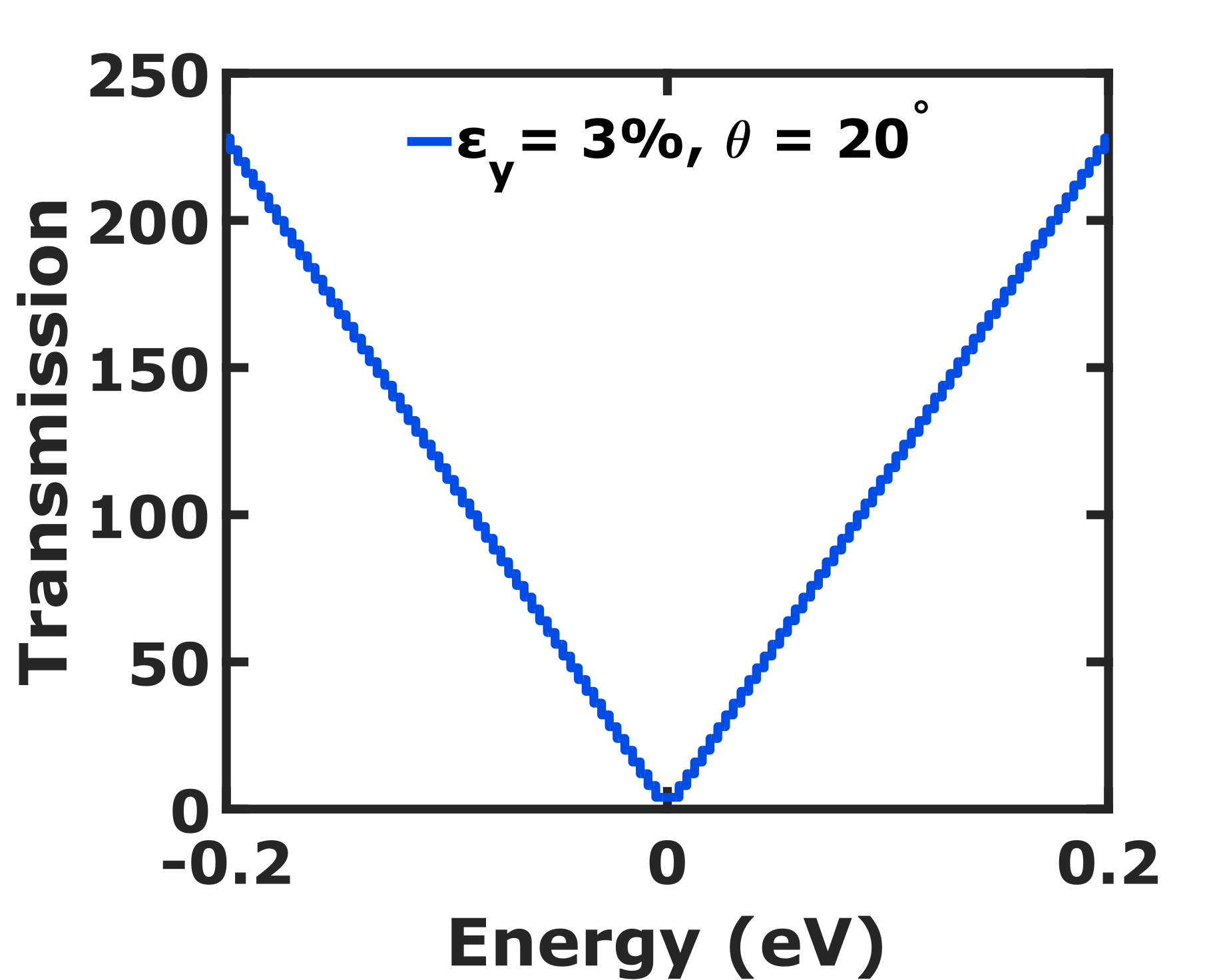}\label{P02_T_3_20}}
	\quad
	\subfigure[]{\includegraphics[height=0.21\textwidth,width=0.229\textwidth]{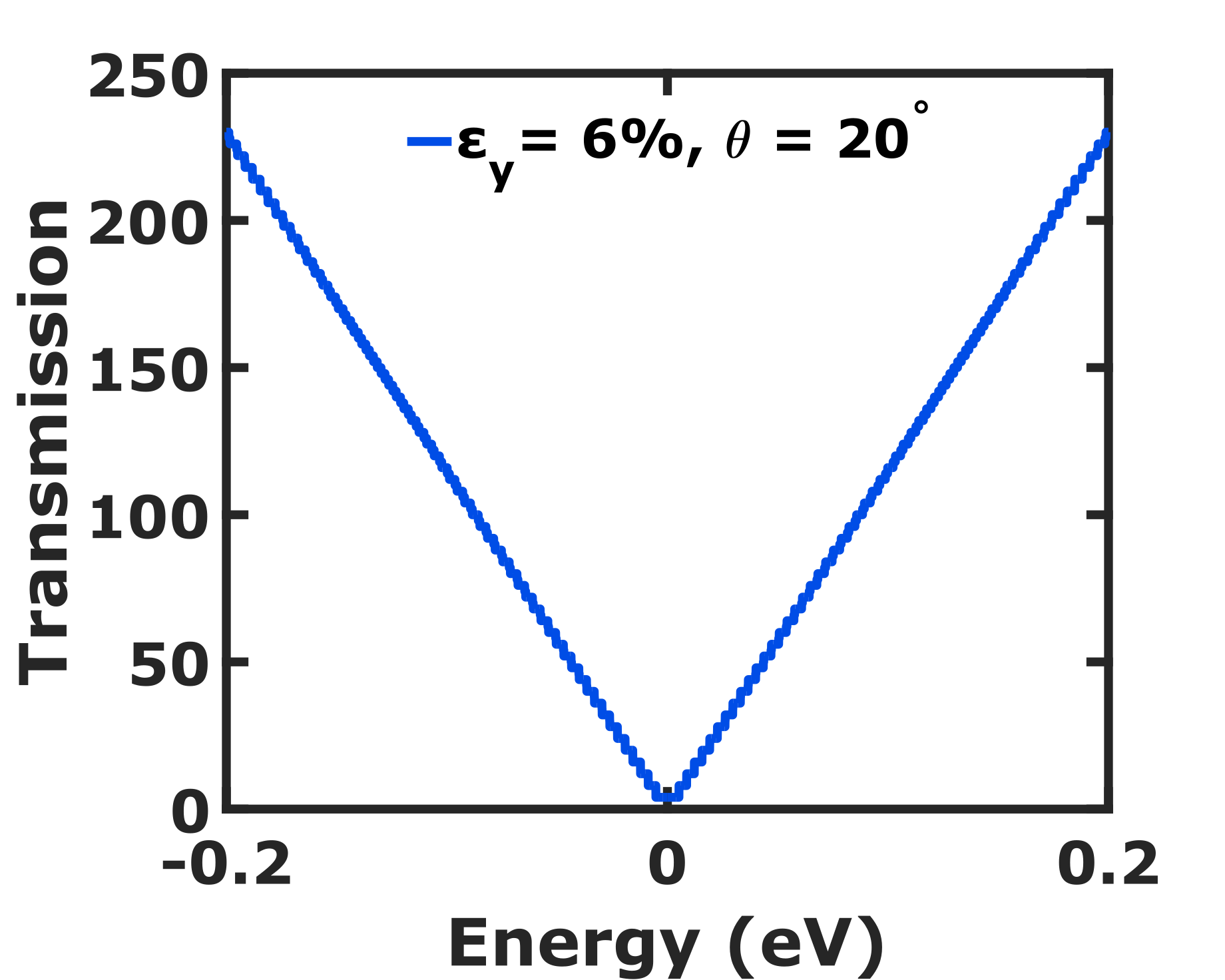}\label{P02_T_6_20}}
	\quad
	\subfigure[]{\includegraphics[height=0.21\textwidth,width=0.229\textwidth]{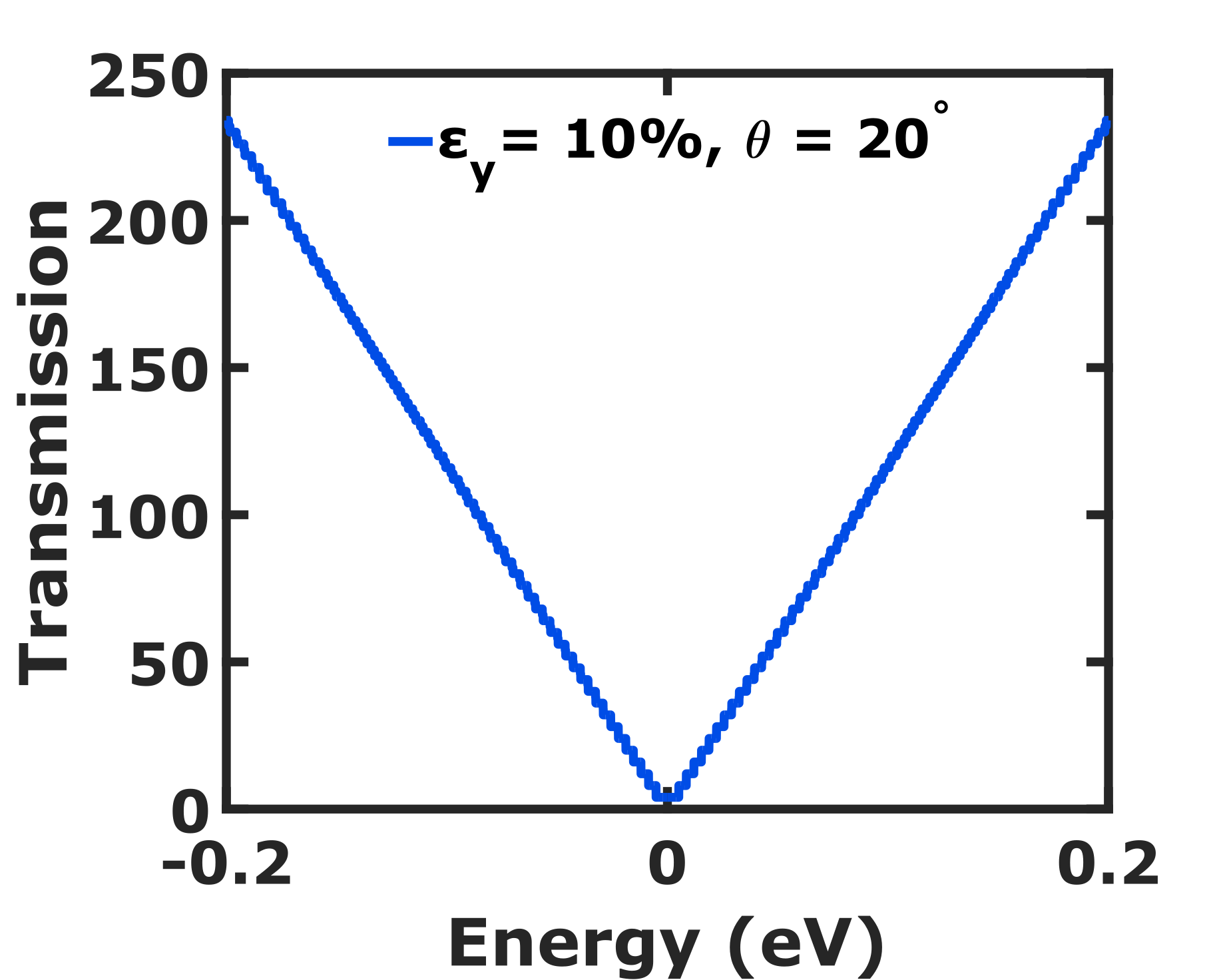}\label{P02_T_10_20}}
	\caption{Variation of transmission as a function of energy within the Fermi-window at different strains $\varepsilon_y$ keeping the transport angle $\theta$ fixed at $20^{\circ}$.}  
	\label{P02_transmission_ballistic}
\end{figure}
In this section, we obtain the AGF of graphene using the theoretical models discussed earlier and obtain a suitable mathematical fit. We discuss the physics behind the predicted results along with their applications and prospects.\\

\begin{figure}
	\centering
	\subfigure[]{\includegraphics[height=0.18\textwidth,width=0.225\textwidth]{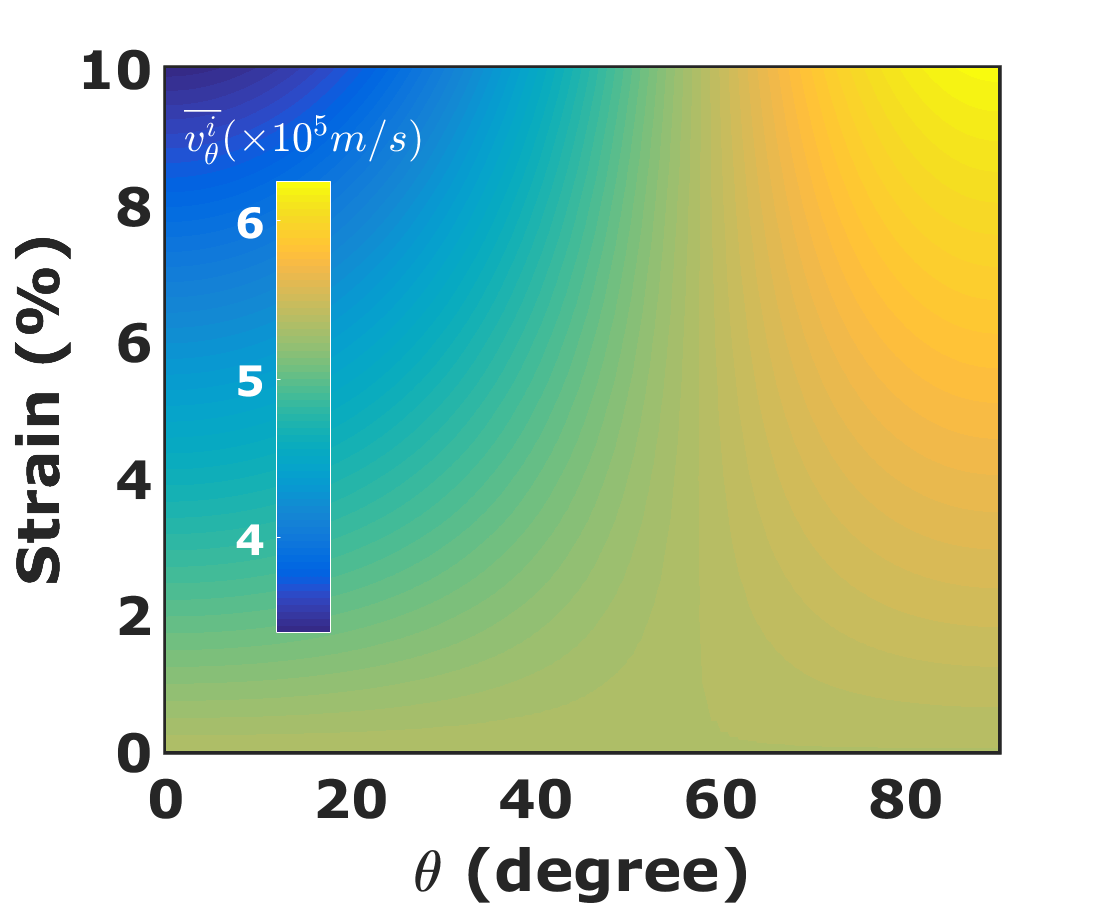}\label{P02_avg_fermi_velocity}}
	\quad
	\subfigure[]{\includegraphics[height=0.18\textwidth,width=0.225\textwidth]{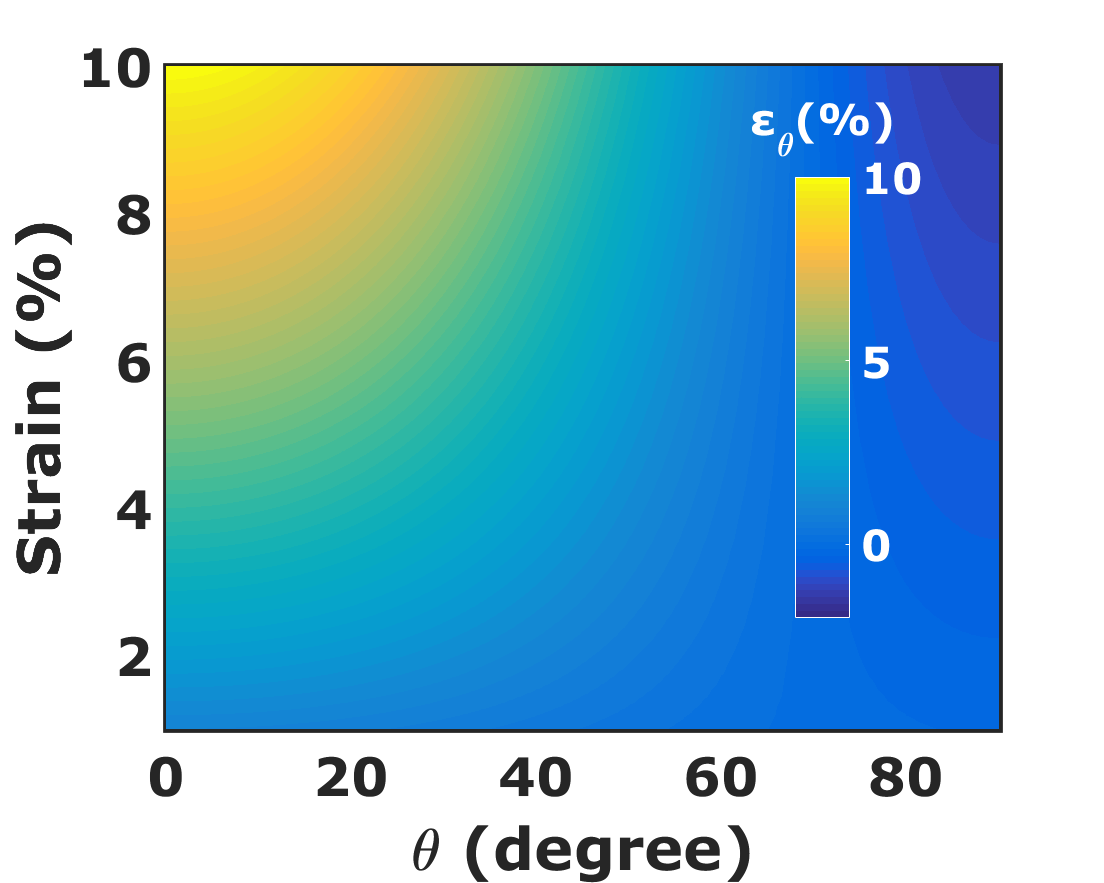}\label{P02_epsilon_l}}	
	\subfigure[]{\includegraphics[height=0.18\textwidth,width=0.225\textwidth]{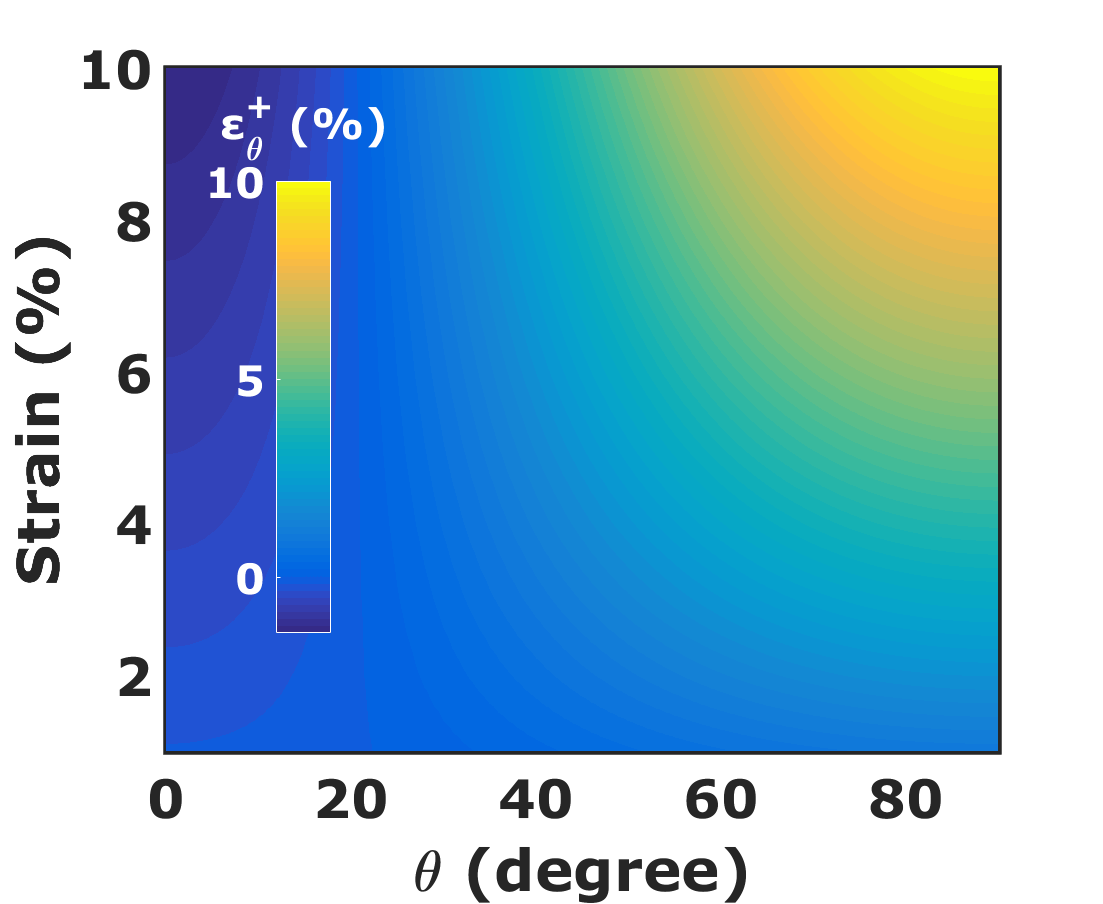}\label{P02_epsilon_t}}	
	\subfigure[]{\includegraphics[height=0.18\textwidth,width=0.225\textwidth]{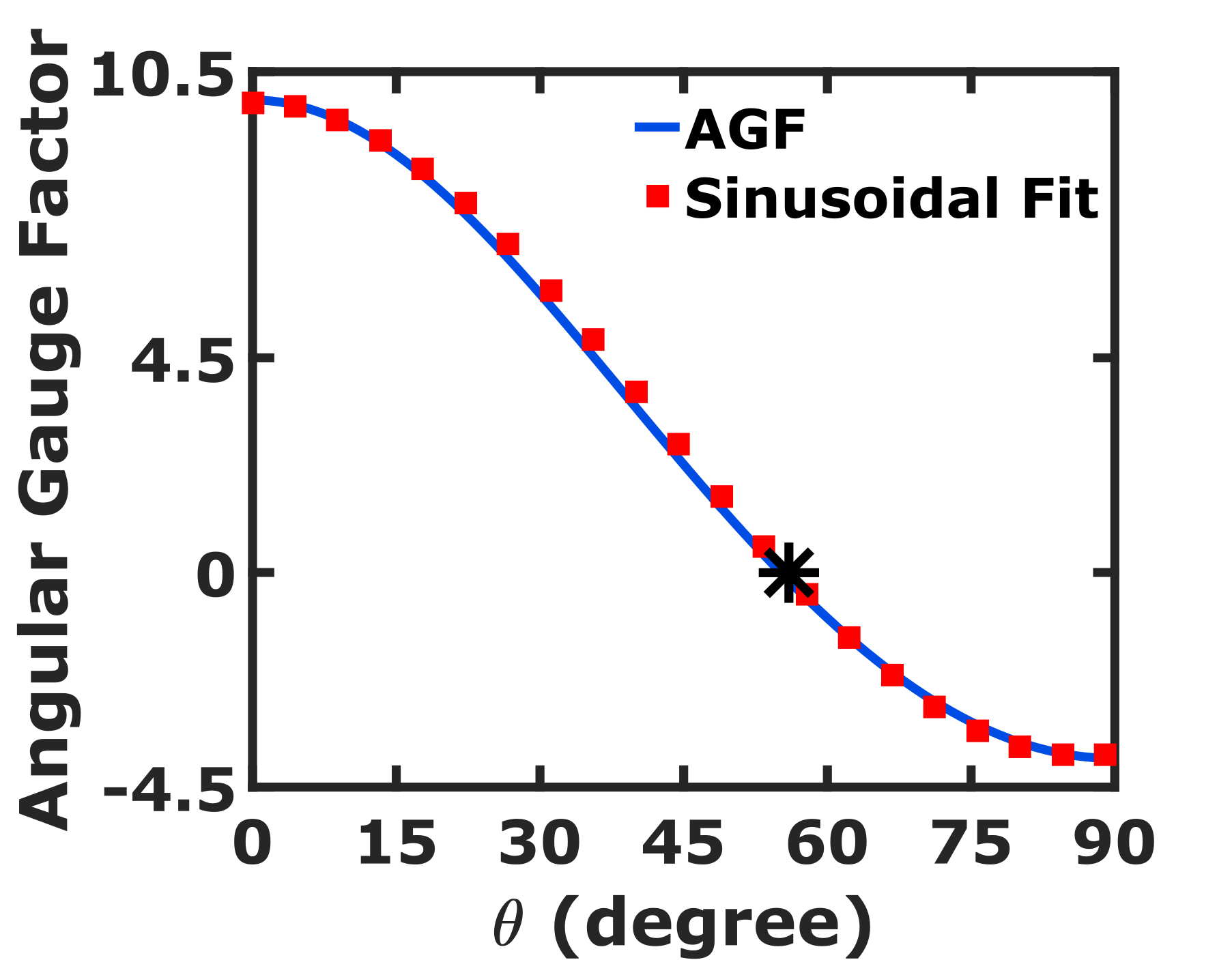}\label{P02_AGF_diffusive}}		\caption{Study of the parameters determining the AGF in diffusive regime.  (a) Color map of the average Fermi-velocity as a joint function of strain ($\varepsilon_y$) and transport angle ($\theta$). Color map of strain along  (b) the transport direction and (c) the transverse direction as a function of $\epsilon_y$ and $\theta$, respectively.(d) Depicts the AGF and its sinusoidal fit as a function of $\theta$.}	
	\label{P02_transport_diffusive}
\end{figure}

\indent We show in Fig.~\ref{P02_Resistance}, the variation of resistance of a 1-$\mu m$ wide ballistic graphene sheet with $\varepsilon_y$ and $\theta$. The resistance at $0\%$ strain is constant irrespective of the direction $\theta$. Nevertheless, the variation in resistance increases with $\theta$ as $\varepsilon_y$ increases. Thus, Fig.~\ref{P02_Resistance} validates the fact that graphene is electrically isotropic at $0\%$ strain and becomes anisotropic on the application of tensile strain~\cite{Pereira2009, Sinha2019}. The anisotropy increases with an increase in strain. The resistance along $\theta=0^{\circ}$ increases by a small amount with an increase in strain. However, it decreases significantly along $\theta=90^{\circ}$ with an increase in strain. We note that the resistance remains constant with applied strain at $\theta = 20^{\circ}$. We show in Fig.~\ref{P02_AGF_ballistic}, the variation of AGF with $\theta$ and the corresponding sinusoidal fit which is a sinusoidal function of the form
\begin{equation}
AGF= A\cos{2\theta}+B,
\label{P02_agf_sin_ballistic}
\end{equation}
where A = 1.475 and B = -0.975. We see in Fig.~\ref{P02_AGF_ballistic} that the AGF along  $0^{\circ}$, $20^{\circ}$, and $90^{\circ}$ are 0.6, 0, and -2.5, respectively. The observed pattern of AGF in the ballistic regime is a result of the deformation of Dirac cone and change in separation of TMs due to strain. The values of AGF at terminal angles $0^{\circ}$ and $90^{\circ}$ are consistent with the longitudinal and transverse GF of graphene obtained in Ref.~\cite{Sinha2019}. The transmissions along $\theta=20^{\circ}$ at different strains are identical as shown in Fig.~\ref{P02_transmission_ballistic}, thereby substantiating our claim of the  resistance invariance direction along $\theta=20^{\circ}$.\\
\indent We show in Fig.~\ref{P02_avg_fermi_velocity}, the variation of average Fermi-velocity with $\theta$ and $\varepsilon_y$.  The average Fermi-velocity is constant at $0\%$ strain along differnt directions `$\theta$' and has a magnitude of $ 5.35\times10^5 m/s $. The average Fermi-velocity decreases sharply along $\theta=0^{\circ}$, becomes zero $\theta\approx60^{\circ}$ and finally slightly increases along $90^{\circ}$ with the increase in strain. The variation in average Fermi-velocity with $\varepsilon_y$ and $\theta$ is similar to the variation of Fermi-velocity (see Table~\ref{table:P02_FV}). Figures~\ref{P02_epsilon_l} and \ref{P02_epsilon_t} depict  $\varepsilon_\theta$ and $\varepsilon^{\dagger}_{\theta}$ as a function of $\varepsilon_y$ and $\theta$. The color maps of $\varepsilon_{\theta}$ and $\varepsilon^{\dagger}_{\theta}$ are mirror image of each other due to two-fold symmetry and isotropic nature of a uniaxially strained graphene (explained in the next paragraph).\\
\indent  Figure~\ref{P02_AGF_diffusive} presents the AGF in diffusive graphene. The plot of AGF with $\theta$ can be approximated by a sinusoidal curve given by
\begin{equation}
AGF= C\cos{2\theta}+D,
\label{P02_agf_sin_diffusive}
\end{equation} 
where C = 6.85 and D = 3. The value of AGF varies sinusoidally between 9.85 to -3.85 and is zero at $56^{\circ}$ in the diffusive regime. We see that ballistic graphene has higher GF along $\theta=90^{\circ}$ whereas diffusive graphene has higher GF along $\theta=0^{\circ}$.\\
\indent Graphene has a six-fold symmetry, but due to the application of tensile strain, its symmetry reduces into a two-fold symmetry.  In Figs.~\ref{P02_AGF_ballistic} and \ref{P02_AGF_diffusive}, the AGF plots have a periodicity of `$\pi$' which is due to the two-fold symmetry of uniaxially strained graphene lattice.\\
\indent The results obtained in this paper are for a strain applied along the zigzag direction. Nevertheless, the results are same for strain along the armchair direction due to the isotropic Poisson's ratio ~\cite{Pereira2009} and identical deformation of the Dirac cone for strain along with armchair and zigzag directions~\cite{Sinha2019}. The explanation for sinusoidal AGF and the application of these results are discussed in the subsequent sections of this paper.\\
\subsection{Physics of sinusoidal AGF}
\begin{figure}	
	\subfigure[]{\includegraphics[height=0.18\textwidth,width=0.225\textwidth]{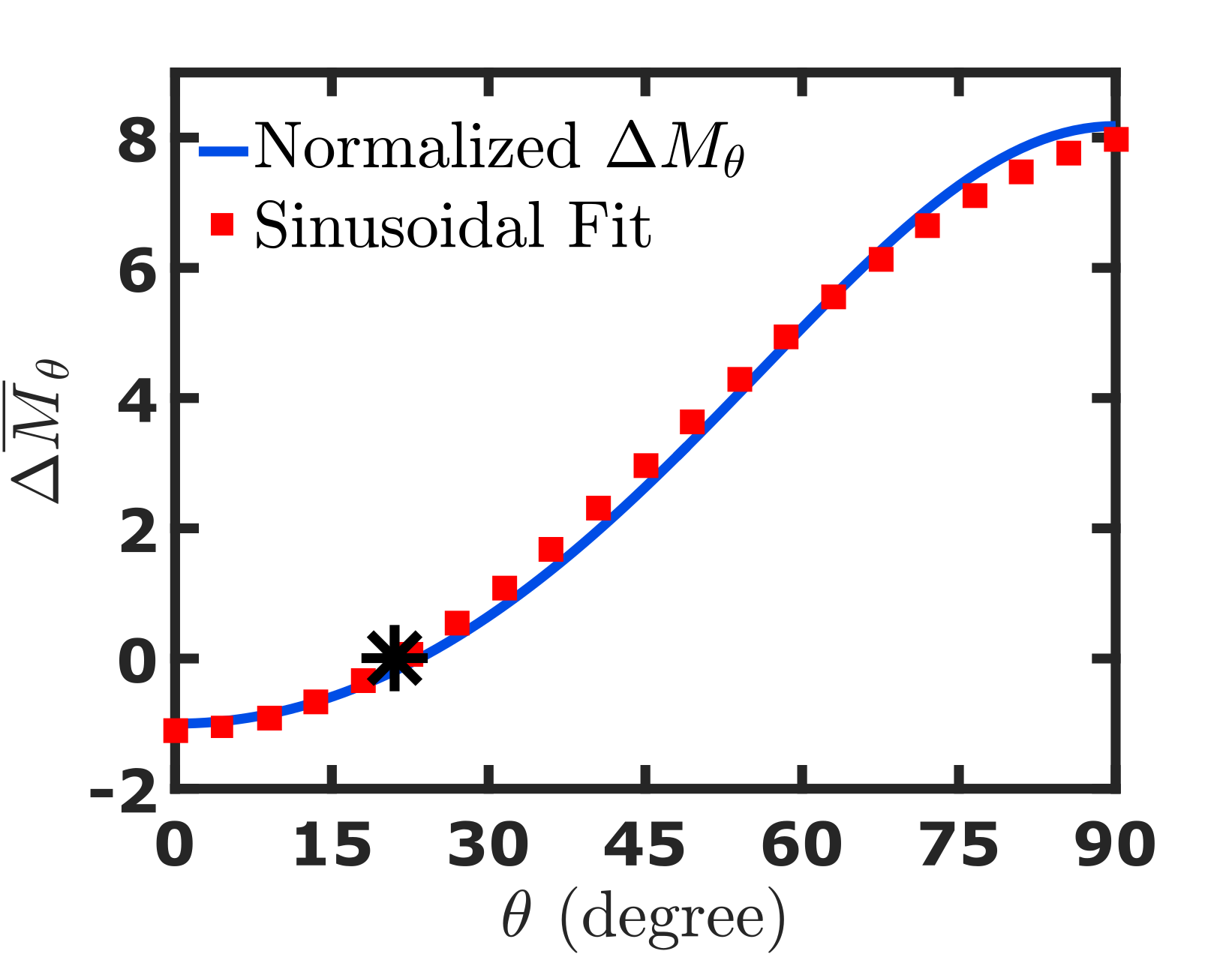}\label{P02_norm_del_modes}}
	\quad
	\subfigure[]{\includegraphics[height=0.18\textwidth,width=0.225\textwidth]{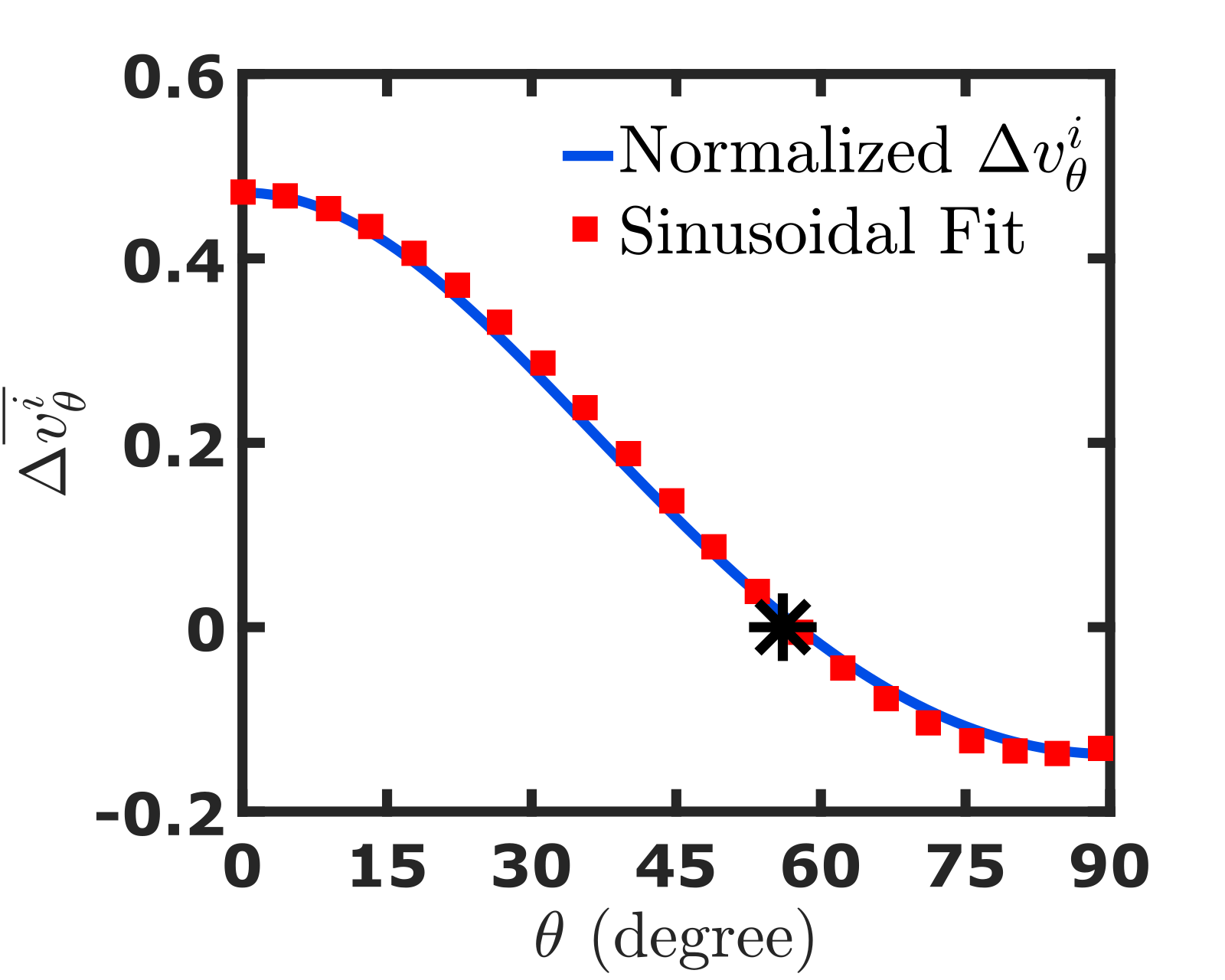} \label{P02_norm_avg_fermi_velocity}}
	\caption{Variation of the (a) normalised mode density (in case of ballistic graphene) and (b) average Fermi velocity (in case of diffusive graphene) as a function of transport direction ($\theta$) and their sinusoidal fit. (a) and (b) provide explanation for the sinusoidal variation of AGF with $\theta$. }
	\label{P02_norm_modes_&_fermi_velocity}
\end{figure}

The mode density along a transport angle $\theta$ changes as a result of the applied strain $\varepsilon_y$ due to the deformation of Dirac cone~\cite{Sinha2019}. The change in mode density is inversely proportional to the change in normalized resistance along the transport angle $\theta$. In other words, AGF is inversely proportional to the change in normalized mode density. The mode density and change in the normalized mode density averaged over the entire strain range [$0\%-10\%$] is mathematically expressed as
\begin{subequations}
\begin{align}  
M_{\theta}(E)=\frac{{w_{\theta}}*L_{\theta+90^{\circ}}(E)}{2\pi}\label{P02_avg_mode_density}, \\
\frac{{\Delta M_{\theta}}(E)}{M_{\theta}(E)}= \frac{\Delta w_{\theta}}{w_{\theta}} + \frac{\Delta L_{\theta +90^{\circ}}(E)}{L_{\theta+90^{\circ}}(E)}. \label{P02_avg_del_mode_density}
\end{align}
\label{P02_avg_modes}
\end{subequations}
where $M_{\theta}(E)$, $w_{\theta}$ and $L_{\theta+90^{\circ}}(E)$ are respectively the mode density, width of the graphene sheet and length of the axis of Dirac cone along the transverse direction  $90^{\circ}+\theta$ as shown in Fig.~\ref{P02_DC_fermi_velocity}. We show in Fig.~\ref{P02_norm_del_modes}, the plot of $\Delta \overline M_{\theta}$ represented by Eq.~\ref{P02_avg_del_mode_density}. Figure~\ref{P02_norm_del_modes} is similar to the reciprocal of AGF in Fig.~\ref{P02_AGF_ballistic}. Thus, the sinusoidal nature of AGF in ballistic graphene is due to the sinusoidal variation of mode density along different $\theta$.\\

\indent The AGF in diffusive graphene depends on the average Fermi-velocity along $\theta$, strain along the transport direction($\varepsilon_{\theta}$) and the transverse direction ($\varepsilon^{\dagger}_{\theta}$). Table~\ref{table:P02_rel_fv_del_l_k} shows the variation of these parameters at transport angles $0^{\circ}$, $45^{\circ}$ and $90^{\circ}$. From the table, we infer that among these parameters, the contribution of average Fermi-velocity in the AGF supersedes all other parameters. We show in Fig.~\ref{P02_norm_avg_fermi_velocity}, the variation of average Fermi-velocity with $\theta$ is similar to a sinusoidal function and resembles the AGF in diffusive regime. Thus, the sinusoidal variation of AGF with transport angle $\theta$ is due to the sinusoidal variation of average Fermi-velocity with transport angle $\theta$.

\begin{table}
	\caption{ Variation of normalised average Fermi-velocity, strain along the transport direction $\theta$ and width along the transverse direction ($\theta + 90^{\circ}$).}
	\centering
	\begin{tabular}{c@{\hskip 1.5 cm} c@{\hskip 1.5 cm} c@{\hskip 1.5 cm} c} 
		
		\hline
		\hline		
		\large$\bm{\theta}$ & \large$\bm {-2\Delta \overline{v}_{\theta}}$ &\large$\bm{\Delta \overline{L}_{\theta}}$ & \large$\bm{-\Delta \overline{w}_{\theta}}$   \\
		
		\hline				
		$\bm{0^{\circ}}$  & 0.52     &  0.055    & 0.008  \\
		$\bm{45^{\circ}}$ & 0.13     &  0.024    & -0.024   \\
		$\bm{90^{\circ}}$ &-0.15     &  -0.008    & -0.055   \\
		\hline
		\hline	
	\end{tabular}	
	\label{table:P02_rel_fv_del_l_k}
\end{table}
\subsection{Application and future scope}
\begin{figure}	
	\subfigure[]{\includegraphics[height=0.18\textwidth,width=0.18\textwidth]{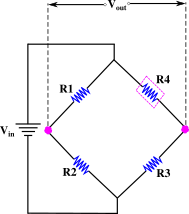} \label{P02_Wheatstone}}
	\quad
	\subfigure[]{\includegraphics[height=0.18\textwidth,width=0.225\textwidth]{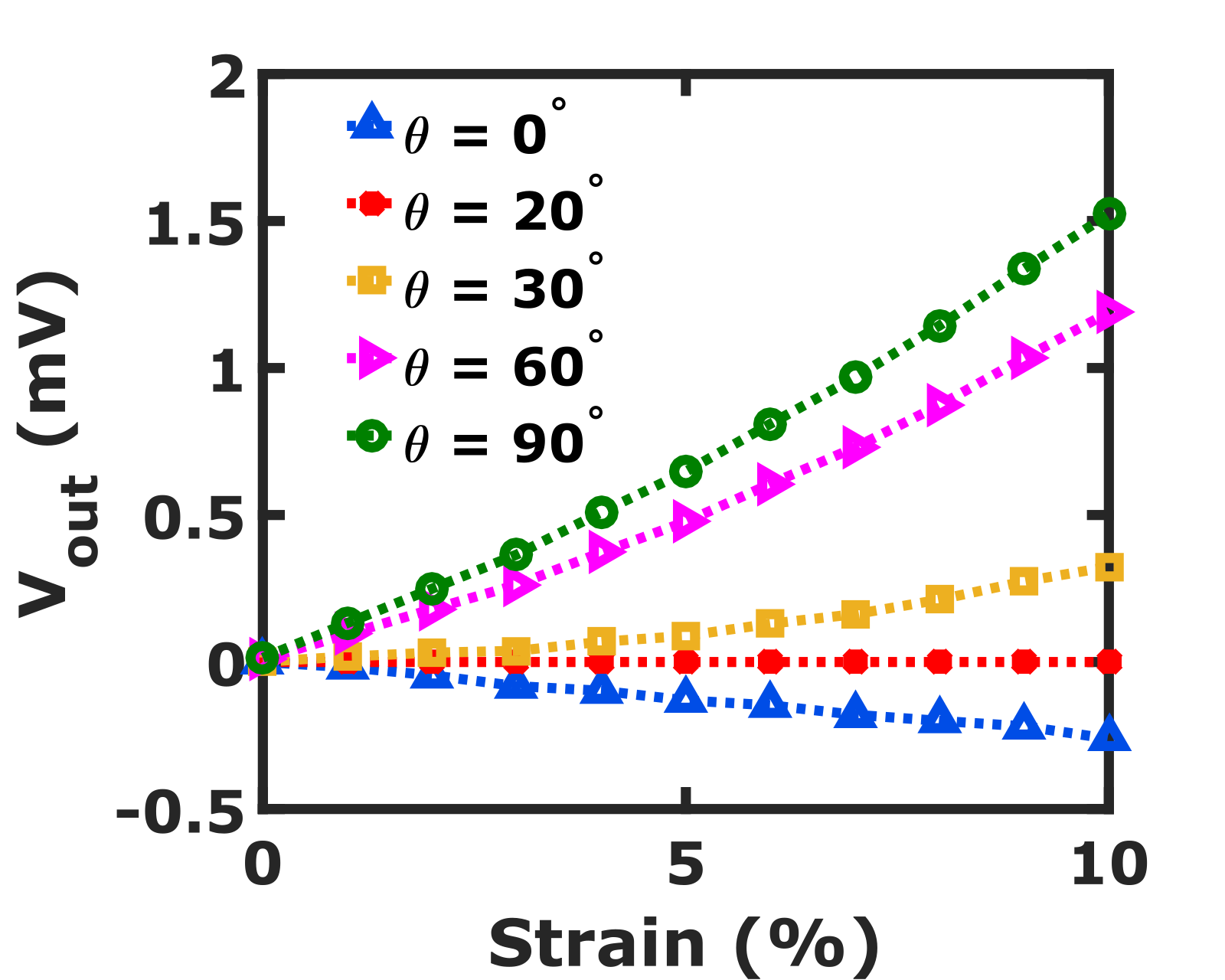}\label{P02_Vout}}	
	\caption{Setup for piezoresistance sensing of a graphene based ballistic nano-sensor using the (a) Wheatstone bridge read-out technique. $R_3$ and $R_4$ are the reference and the strain sensors respectively. (b) Variation in the output voltage ($V_{out}$) of the wheatstone bridge with strain ($\varepsilon_y$) and transport direction ($\theta$). $V_{out}$ varies linearly with the strain along different directions.}  
	\label{P02_Wheatstone_Vout}
\end{figure}

Piezoresistance sensing is commonly done using the Wheatstone bridge readout technique. We show in Fig.~\ref{P02_Wheatstone}, a Wheatstone bridge based piezoresistance sensing setup for ballistic graphene. The Wheatstone bridge consists of identical ballistic graphene resistors $R_1$,$R_2$,$R_3$ and $R_4$. At zero strain, the resistors are equal, and the bridge is in a balanced state. $R_4$ is the strain sensor, and the other resistors act as reference resistors. When subjected to strain, the resistance of $R_4$ changes, which results in the generation of a potential difference ($V_{out}$), as shown in Fig.~\ref{P02_Vout}. Figure~ \ref{P02_Vout} shows $V_{out}$ versus strain for different values of $\theta$ when the input voltage $V_{in}$ is maintained at 0.02 eV. $V_{out}$ is highest when $\theta=90^{\circ}$. In addition, $V_{out}$ shows a linear dependence with strain at different $\theta$. Thus, ballistic graphene can act as a strain sensor and shows maximum sensitivity at $\theta=90^{\circ}$. Ballistic graphene can be used to detect explosives, gases, etc. provided the strain generated is mapped with the vapor density of explosives or gases.~\cite{Boisen2011, Seena2011, Nelson2006, Pinnaduwage2004}. \\
\indent A major problem pertaining to the strain sensor is the presence of an inherent residual strain. Consequently, $V_{out}$ is non-zero even when no external strain is applied. In fact, during explosive detection or chemical sensing, the reference sensors are vulnerable to unwanted strain by the vapors of explosives and gases~\cite{Boisen2011, Seena2011, Nelson2006, Pinnaduwage2004}. In this paper, we propose a solution to overcome this problem without removing the residual or unwanted strain. As discussed earlier, the AGF of ballistic and diffusive graphene is zero at the critical angles. In this direction, the resistance does not change even in the presence of strain. The resistance along the critical angle is always equal to the resistance along any other direction $\theta$ at $0\%$ strain. The above discussion is equally valid for diffusive graphene also. Hence, we propose a graphene-based reference piezoresistor with the transport along the critical angle in the Wheatstone bridge read-out technique.\\
\indent Due to the excellent electro-mechanical properties, graphene is also a strong contender for materials used in flexible electronics~\cite{Kim2009, Sharma2013}. The critical angle may find application in future flexible devices where a constant current is required despite the presence of a variable strain. \\
\indent The existence of a critical angle in ballistic graphene is due to the unique deformation pattern of Dirac cone as a result of applied uniaxial strain. Similar studies for 2-D Dirac materials such as Silicene and Germanene etc. in ballistic regimes can be undertaken for similar applications in the future.
   
\section{Conclusion} \label{section4}

In summary, we investigated the angular gauge factor of graphene in the ballistic and diffusive regimes using highly efficient quantum transport models. It was shown that the angular gauge factor in both ballistic and diffusive graphene between $0^{\circ}$ to $90^{\circ}$ bears a sinusoidal relation with a periodicity of $\pi$ due to the reduction of six-fold symmetry into a two-fold symmetry as a result of applied strain. The angular gauge factor is zero at critical angles $20^{\circ}$ and $56^{\circ}$ in ballistic and diffusive regimes, respectively. Based on these findings, we propose a graphene-based ballistic nano-sensor, which can be used as a reference piezoresistor in a Wheatstone bridge read-out technique. The reference sensors proposed here are unsusceptible to inherent residual strain present in strain sensors and unwanted strain generated by the vapors in explosives detection. The theoretical models developed in this paper can be applied to explore similar applications in other 2D-Dirac materials. The proposals made here potentially pave the way for implementation of NEMS strain sensors based on the principle of ballistic transport, which will eventually replace MEMS piezoresistance sensors with a decrease in feature size. The presence of strain insensitive ``critical angle'' in graphene may be useful in flexible wearable electronics also.
    
\begin{acknowledgments}

The Research and Development work undertaken in the project under the Visvesvaraya Ph.D. Scheme of Ministry of Electronics and Information Technology, Government of India, is implemented by Digital India Corporation (formerly Media
Lab Asia). This work was also supported by the Science and
Engineering Research Board (SERB), Government of
India, Grant No. EMR/2017/002853 and Grant No. STR/2019/000030 and the Ministry of Human Resource and Development (MHRD), Government of India, Grant no. STARS/APR2019/NS/226/FS under the STARS scheme. The author Abhinaba Sinha was also supported by NNETRA project under Sanction Order: DST/NM/NNetRA/2018(G)-IIT-B at IIT Bombay.

\end{acknowledgments}
\appendix 

\section{{Separation between adjacent transverse modes} \label{P02_app1}}
The separation between adjacent $k-$ states of conduction electrons in reciprocal space is determined by periodic boundary condition~\cite{Ashcroft}. Electrons in conduction band behave as nearly free electrons. The wave-function of these electrons are expressed as  
\begin{equation}
\psi= Ae^{i k\cdot r}.
\label{P02_app1_eqpw}
\end{equation}
Equation~\eqref{P02_app1_eqpw} represents a plane-wave equation travelling along $r$. Let us consider a crystal of length L and wave-function at r=0 and r=L be denoted by $\psi_{0}$ and $\psi_{L}$ respectively. The wave-functions at the boundaries are equal as a result of periodic boundary condition. Thus,
\begin{equation}
\psi_{0}=\psi_{L}.
\label{P02_app1_eqpbc}
\end{equation}
Using Eqs.~\eqref{P02_app1_eqpw} and \eqref{P02_app1_eqpbc}, we obtain the allowed $k-$states which is expressed as
\begin{equation}
k=\frac{2n\pi}{L},
\label{P02_app1_eqkstates}
\end{equation}
where n=0,1,2,3.....
The separation between $k$-states in reciprocal space is $2\pi/L$, where L is the length of the crystal. Extending the same reasoning along the width (w), we obtain a separation of $2\pi/w$ between adjacent transverse modes (TMs) along the width in reciprocal space.\\

\section{{Average Fermi-velocity under Dirac cone approximation} \label{P02_app2}}
Under the Dirac cone approximation, the velocity of electrons along $\phi$ at different energies are equal to the Fermi-velocity~(see Fig.~\ref{P02_Fermi-velocity}). The magnitude of Fermi-velocity along different directions are equal in unstrained graphene. Application of a tensile strain ($\varepsilon_y$) results in anisotropic Fermi-velocity~(see Fig.~\ref{P02_fermi_velocity_10}). As a result, evaluation of the conductivity of strained graphene using Eq.~\eqref{P02_eqc} will require the value of average Fermi-velocity along $\theta$. \\
\indent Figure~\ref{P02_DC_fermi_velocity} illustrates the methodology for calculating average Fermi-velocity. The direction of Fermi velocity($\phi$) forming angle in the range $-\frac{\pi}{2}$ to $\frac{\pi}{2}$ with respect to $\theta$ have velocity components along $\theta$. The Fermi-velocity ($v^{i}_{\phi}$) along $\phi$ is expressed as
\begin{equation}
v^{i}_{\phi}=\frac{1}{\hbar}\{\nabla E^{i}(k)\}\bigg|_{k=k^{i}_{\phi}}.
\end{equation}
The component of $v^{i}_{\phi}$ along $\theta$ is $v^{i}_{\phi}\cdot k^{i}_{\theta}$, where $\hat{k^{i}_{\theta}}$ is a unit-vector along $\theta$. Thus, the mean of these components for different values of $\phi$ along $\theta$ is given by 
\begin{equation}
\overline{v^{i}_{\theta}}=\frac{1}{\pi} \int_{-\frac{\pi}{2}}^{\frac{\pi}{2}} \{v_{\phi}^{i}\cdot \hat{k^{i}_{\theta}}\} d\phi.
\end{equation}

\bibliographystyle{apsrev4-1}
\bibliography{reference01}

\end{document}